\documentclass[10pt, conference]{IEEEtran}   

\usepackage{amsmath,amsfonts}
\usepackage{textcomp}

\usepackage{algorithm}
\usepackage[noend]{algpseudocode}
\usepackage[noend]{algpseudocode}
\algrenewcommand\algorithmicrequire{\textbf{Input:}}
\algrenewcommand\algorithmicensure{\textbf{Output:}}
\algrenewcommand{\algorithmiccomment}[1]{\hfill{\color{blue}$\triangleright$#1}}

\usepackage{float}
\floatstyle{ruled}

\usepackage{cite}
\usepackage[hidelinks]{hyperref}
\hypersetup{
    pdfcreator={},%
    pdfproducer={},%
    pdfsubject={Accepted version to IEEE INFOCOM 2024},%
    pdftitle={A Practical Near Optimal Deployment of Service Function Chains in Edge-to-Cloud Networks},%
    pdfauthor={R. Behravesh, D. Breitgand, D. H. Lorenz, and D. Raz}
}
\usepackage[capitalise]{cleveref}
\usepackage{url}

\usepackage[caption=false,font=footnotesize]{subfig}
\usepackage{stfloats}
\usepackage{graphicx}
\usepackage{booktabs}  
\usepackage{tabularx}  
\usepackage{multirow}
\usepackage{array}

\usepackage{verbatim}
\usepackage[normalem]{ulem}
\usepackage{xargs} 
\usepackage{xcolor}
\usepackage{xspace}
\xspaceaddexceptions{]\}}
\usepackage{mathtools}
\usepackage[inline]{enumitem}  

\newcommand{\Small}{\footnotesize} 

\newcommand{\Shortstack}[2][t]{\begin{tabular}[#1]{@{}l@{}}#2\end{tabular}}

\usepackage{ifdraft}
\ifoptionfinal{%
\usepackage[final]{fixme}
}{%
\usepackage[draft]{fixme}

}
\fxsetup{mode=multiuser,theme=color,inline,nomargin} 
\FXRegisterAuthor{DL}{dl}{\color{blue}\fbox{DL}}
\FXRegisterAuthor{DB}{db}{\color{green!50!black}\fbox{DB}}
\FXRegisterAuthor{RB}{rb}{\color{teal!70!black}\fbox{RB}}
\FXRegisterAuthor{DR}{dr}{\color{violet}\fbox{DR}}

\newcommand{\DL}[1]{\DLnote{#1}}
\newcommand{\DB}[1]{\DBnote{#1}}
\newcommand{\RB}[1]{\RBnote{#1}}

\newcommand{\capacity}[1]{\operatorname{cap}(#1)}
\newcommand{\cost}[1]{\operatorname{cost}(#1)}
\newcommand{\dem}[1]{\operatorname{dem}(#1)}

\newcommand{\partreject}{\mathit{RejectedByRounding}}

\newcommand{\dc}{DC\xspace}
\newcommand{\dcs}{DCs\xspace}
\newcommand{\set}[1]{\{#1\}}
\newcommand{\Set}[1]{\mathbf{#1}}

\newcommand{\ue}{\textsc{u}}
\newcommand{\term}{\textsc{t}}

\newcommand{\dfict}{d_\Phi}

\newcommand{\dflow}[4][a]{\chi^{#1,(#3)}_{#2,(#4)}}
\newcommand{\tflow}[4][a]{\overline\chi^{#1,(#3)}_{#2,(#4)}}

\newcommand{\pst}[1][st]{\Set p_{#1}}
\newcommand{\Pst}[1][st]{\Set P_{\!#1}}

\newcommand{\appv}[3][a]{z^{#1,#2}_{#3}}
\newcommand{\appe}[5][a]{w^{#1,(#2)}_{#4,(#3), #5}}
\newcommand{\appbw}[3][a]{w^{#1,(#2)}_{(#3)}}
\newcommand{\appp}[3][a]{w^{#1,(#2)}_{\Set p_{#3}}}
\newcommand{\userv}[3][u]{x^{#1, #2}_{#3}}

\newcommand{\userp}[3][u]{y^{#1,(#2)}_{\pst[#3]}}

\newcommand{\Adu}{\textsc{ADU}\xspace}
\newcommand{\adu}{\textsc{adu}\xspace}
\newcommand{\adus}{\textsc{adu}s\xspace}

\newcommand{\Ecu}{\textsc{ECU}\xspace}
\newcommand{\ecu}{\textsc{ecu}\xspace}
\newcommand{\ecus}{\textsc{ecu}s\xspace}

\newcommand{\Bwu}{\textsc{BWU}\xspace}
\newcommand{\bwu}{\textsc{bwu}\xspace}
\newcommand{\bwus}{\textsc{bwu}s\xspace}

\newcommand{\solution}{\boldsymbol{\Psi}}
\newcommand{\fraqopt}{\boldsymbol{\chi}}

\newcommand{\preasonable}[3][t]{\rho_{#2#1}(#3)} 
\newcommand{\Preasonable}[2][t]{\Pst(\rho_{#2#1})} 
\DeclarePairedDelimiter\abs{\lvert}{\rvert}
\DeclarePairedDelimiter\norm{\lVert}{\rVert}

\newcommand{\ignore}[1]{} 

\newcommand{\blue}[2][blue]{{\color{#1}#2}}

\newtheorem{define}{Definition}
\newtheorem{theorem}{Theorem}

\Crefname{define}{Definition}{Definitions}

\newfloat{formulation}{tbp}{lop}
\floatname{formulation}{Formulation}
\newfloat{problem}{tbp}{lop}
\floatname{problem}{Problem}

\Crefname{problem}{Problem}{Problems}
\crefname{algorithm}{Alg.}{Algs.}

\crefname{formulation}{}{Formulations}
\Crefname{formulation}{Formulation}{Formulations}

\crefname{prob}{}{}
\creflabelformat{prob}{#2\textbf{\textsc{Sfc-dp}}#3}

\crefname{milp}{}{}
\crefname{milp}{}{Formulations}
\Crefname{milp}{Formulation}{Formulations}
\creflabelformat{milp}{#2\textbf{#1}#3}   

\crefname{flow}{}{}
\crefname{flow}{}{Formulations}
\Crefname{flow}{Formulation}{Formulations}
\creflabelformat{flow}{#2\textbf{#1}#3}   

\crefname{pranos}{}{}
\crefname{pranos}{Alg.}{Algs.}
\Crefname{pranos}{Algorithm}{Algorithms}
\creflabelformat{pranos}{#2\textbf{#1}#3}   

\setlist[description]{leftmargin=0.6in}
\setlist[itemize]{leftmargin=*}
\newlist{enumproblem}{enumerate}{1}
\setlist[enumproblem]{label=(O\arabic*), align=left, leftmargin=\parindent, labelwidth=!}
\crefname{enumproblemi}{}{}

\crefname{equation}{}{Eqs.}
\Crefname{equation}{Equation}{Equations}

\newcommand{\pranos}{\textsc{Pranos}\xspace}
\begin{document}

\title{A Practical Near Optimal Deployment of Service Function Chains in Edge-to-Cloud Networks}
\author{}
\author{Rasoul Behravesh, David Breitgand, Dean H. Lorenz, and Danny Raz}%



\makeatletter
\def\ps@IEEEtitlepagestyle{%
  \def\@oddfoot{\mycopyrightnotice}%
  \def\@oddhead{\hbox{}\@IEEEheaderstyle\leftmark\hfil\thepage}\relax
  \def\@evenhead{\@IEEEheaderstyle\thepage\hfil\leftmark\hbox{}}\relax
  \def\@evenfoot{}%
}
\def\mycopyrightnotice{%
  \hspace{-.04\textwidth}
  \fbox{\begin{minipage}{1.08\textwidth}
  \centering\scriptsize
  \copyright~2024 IEEE. Personal use of this material is permitted. Permission from IEEE must be
obtained for all other uses, in any current or future media, including reprinting/republishing this material for advertising or promotional purposes, creating new collective works, for resale or redistribution to servers or lists, or reuse of any copyrighted component of this work in other works.
  \end{minipage}}
}
\renewcommand\@oddhead{\hbox{}\@IEEEheaderstyle\leftmark{R.~Behravesh, D.~Breitgand, D.~H.~Lorenz, D.~Raz}\hfil\thepage}\relax
\makeatother
\maketitle

\begin{abstract}
\ignore{
Mobile edge computing offers a plethora of opportunities to develop new applications, offering much better quality of experience to the users. A fundamental problem that has been thoroughly studied in this context is deployment of Service Function Chains (SFCs) into a physical network on the spectrum between edge and cloud. This problem is known to be NP-hard. Because of its practical importance high quality sub-optimal solutions are of great interest.
}
Mobile edge computing offers a myriad of opportunities to innovate and introduce novel applications, thereby enhancing user experiences considerably. A critical issue extensively investigated in this domain is efficient deployment of Service Function Chains (SFCs) across the physical network, spanning from the edge to the cloud. This problem is known to be NP-hard. As a result of its practical importance, there is significant interest in the development of high-quality sub-optimal solutions.

In this paper, we consider this 
problem and propose a novel near-optimal heuristic that is extremely efficient and scalable. We compare our solution to the state-of-the-art heuristics and to the theoretical optimum. In our large scale evaluations, we use realistic topologies which are previously reported in the literature. We demonstrate that the execution time offered by our solution grows slowly as the number of Virtual Network Function (VNF) forwarding graph embedding requests grows, and it handles one million requests in slightly more than $20$ seconds for $100$ nodes and $150$ edges physical topology.\ignore{ \DB{maybe we need to revise this result and use the one for even larger topology}. Furthermore, we demonstrate that our heuristic handles hot spots much better than the state-of-the-art, resulting in much lower request rejection rate.
\RB{ -- Add complete form of "VNF" -- if we designate a name to our algorithm, we can refer to in in the abstract (or even the title can be complete form of the algorithm's name) -- we can give percentage instead of "much lower rejection"}}

\end{abstract}


\def\figscale{0.63}

\widowpenalty=10000
\section{Introduction}\label{sec:intro}


Network Function Virtualization (NFV) and Software Defined Networking (SDN) provide a cost-efficient, flexible, and agile approach to deploy and manage services and applications in the edge-to-cloud spectrum. With this approach, custom physical network appliances become Virtual Network Functions (VNFs), which can efficiently run on off-the-shelf hardware. This dramatically reduces capital expenditures and increases flexibility. In this new paradigm, an application is a chain of VNFs (aka service function chain) expressed by a VNF-Forwarding Graph (VNF-FG). Deploying a Service Function Chain (SFC) is fundamental to network virtualization. Hence its multiple variants were extensively studied~\cite{KAUR2020100298}. 

Deploying SFC requires solving two sub-problems: (1) embedding VNF-FG in the physical network substrate and (2) steering individual sessions traffic through the embedded VNF-FG (i.e., setting forwarding rules in the \emph{embedded} VNF-FG). The first sub-problem implies mapping VNFs on the physical nodes and logical links interconnecting VNFs in VNF-FG onto paths in the physical network substrate at minimal cost. Capacity requirements of VNFs, bandwidth requirements of the links and latency requirements of the links must be respected. This problem is equivalent to the well-studied Virtual Network Embedding Problem (VNEP), which is known to be strongly NP-hard~\cite{Rexford2008, RostSchmidt-ToN2020}. The second sub-problem, i.e., traffic steering through the embedded VNF-FG, is similar to the unsplittable multi-commodity flow problem, where traffic entering the embedded VNF-FG through a given edge node corresponds to a distinct commodity. This is another well-studied problem, which is known to be NP-complete~\cite{EvenIS-Multicommodity-Unsplittable-Flow-1975}.

In this work, we consider a typical business scenario, in which a physical network provider offers applications on top of a physical network. Users make requests for application sessions and each request can either be accepted or rejected. Users can access the physical network through dedicated point of presence edge nodes only. We consider an offline setting, in which the network provider is presented with a set of application requests, where each application corresponds to an SFC. The network provider wishes to accept as many requests as possible at the lowest cost possible. As explained above, this requires embedding the VNF-FGs of the SFCs and steering traffic through the embedded VNF-FGs at minimal cost, while respecting capacity and latency constraints.

One particularly important  
scenario, in which the SFC deployment problem should be solved at extreme scale is 5G/6G mobile network.
Indeed, the Key Performance Indicators (KPIs) envisioned for 5G stipulate scaling to one million connections per square kilometer~\cite{Liu-5GKPI-2020}. Even with the expected overbooking factor of 1/50~\cite{OughtonFGB-Netherlands5G-2019}, 5G networks have to support efficient embedding of 20K simultaneously active sessions per square kilometer. Thus, developing scalable heuristic solutions to SFC embedding is of great practical value. 

Our solution proposes a novel heuristic, PRActical Near Optimal SFC Embedding (\ref{alg:prano}), based on a new LP relaxation, rounding, and handling of latency requirements. We discuss the similarities and differences with \mbox{state-of-the-art} LP-based solutions~\cite{RostSchmid-RandomizedRounding-ToN2019, MunkRostRackeSchmid-relax-2021} in detail in Section~\ref{sec:related}. 
Our specific contributions are as follows.
\begin{itemize}
    \item We formulate SFC deployment using a cost minimization Mixed Integer Linear Programming (MILP) in \cref{sec:problem}
    \item We present a novel heuristic, \ref{alg:prano}, in \cref{sec:solution}. \ref{alg:prano} comprises two parts: (1) an LP relaxation that avoids path enumeration for VNF-FG  embedding sub-problem in Subsection~\ref{sec:flow}, (2) a rounding algorithm that solves the traffic steering problem on the embedded VNF-FG and provides a \emph{deterministic} upper bound on the number of rejected SFC embedding requests in Subsection~\ref{sec:heur3};
    \item We offer a novel heuristic for handling latency constraints while embedding SFCs in Subsection~\ref{sec:lat}; 
    \item We evaluate our proposal via large-scale and extensive simulations in \cref{sec:evaluation}, and demonstrate that it is significantly superior to the state-of-the-art heuristic HEU\_Cost~\cite{harutyunyan2019latency, harutyunyan2020latency} and is near optimal when compared to the theoretical optimum. To the best of our knowledge, this is the largest simulation study reported for the problem so far.
    \item In \cref{sec:related}, we present related works and in \cref{sec:conclusions} we present our conclusion and 
    discuss future work.

    \ignore{Furthermore, using topologies of the size that one can expect in large metropolitan areas, we demonstrate that the execution time of the proposed heuristic stays essentially flat for the topology of a given size and for a given mixture of applications as more users join the network making it scalable to the expected UE density stipulated in the 3GPP 5G standards~\cite{3gpp}. that it scales to large realistic scenarios and provides significant improvement over SOTA. }
\end{itemize}

\section{Network Model and Problem Definition}\label{sec:problem}


\subsection{Substrate Network}\label{subsec:substrate}

We model the substrate network as a weighted undirected graph and denote it by $G(\Set D, \Set E)$, where $\Set D$ is the set of Data Centers (\dcs) comprising the substrate nodes and $\Set E$ is the set of substrate links \cite{ChowdhuryRB-TON2012}.\footnote{We assume that \dcs are interconnected in the physical substrate through the forwarding paths constructed by some reasonable routing protocols. The details of the routing protocols in the physical network are not important for the rest of the discussion.} We denote by $\Pst$ the set of available substrate paths from source node $s$ to target node $t$. The latency of a path $\pst \in \Pst$ is denoted by $L(\pst) = \sum_{e_{mn} \in \pst}L(m,n)$, where $L(m,n)$ denotes the latency of substrate link $e_{mn} \in \Set E$.

Following cloud computing approach, we express resource sizing in Elastic Capacity Units (\Ecu)~\cite{SAP-HANA}. The elastic capacity of a substrate node $d \in \Set D$ is denoted by $\capacity{d}$ and is measured in \ecus.\footnote{By setting the capacity of a \dc to $0$, we can model network forwarding elements, such as switches, that do not host VNFs.} We use Bandwidth Units (\Bwu) to express link resources; the bandwidth capacity of a substrate link $e_{mn} = (m,n) \in \Set E$ is denoted by $\capacity{m, n}$ and is measured in \bwus.

We use a generic \emph{cost} to model the optimization goal. Each \dc may have a different cost per \ecu. We denote the cost of reserving one \ecu on \dc $d$ by $\cost{d}$. Similarly, we denote by $\cost{m,n}$ the cost of reserving one \bwu on $e_{mn}$.

\begin{table}
    \caption{Notations}
    \label{tab:notations}
    \centering\Small
    \begin{tabularx}{\columnwidth}{@{}lXr@{}}
        \toprule
        \textbf{Notation} & \textbf{Description} \\ 
        \midrule
        $d \in \Set D$ &  a substrate \dc (node) 
        & \multirow[b]{2}{*}{\llap{\color{blue}\shortstack[r]{Substrate\\Network}}} \\
        $e_{mn} \in \Set E$ & substrate link between \dcs $m,n$ \\
        $\pst \in \Pst$ & substrate path between \dcs $s$ and $t$ \\
        $\cost{d}$ & cost of elastic capacity on \dc $d$ \\
        $\cost{m,n}$ & cost of elastic bandwidth on $e_{mn}$ \\
        $\capacity{d}$ & size [\ecu] of elastic capacity on $d$\\
        $\capacity{m,n}$ & size of available bandwidth on $e_{mn}$ \\
        $L(m,n), L(\pst)$ & latency of $e_{mn}$, latency of $\pst$ \\
        \midrule
        $a \in \Set A$ & an application 
        & \multirow[b]{3}{*}{\llap{\color{blue}\shortstack[r]{Service\\Function\\Chains}}} \\
        $f^a_i \in \Set F^a$ & a VNF in the chain of application $a$ \\
        $e^a_{ij}$ & application link between functions $f^a_i$ and $f^a_j$ \\
        $f^a_{\ue}$ & root of application $a$ (represents the user location) \\
        $f^a_{\term}$ & terminator for all leaf functions in tree of $a$ \\
        $L(e^a_{ij})$ & maximal allowed latency between $f^a_i$ and $f^a_j$ \\
        \midrule
        $u \in \Set U$ & a user SFC request 
        & \llap{\color{blue}Demand} \\
        $\dem{u}$ & demand [\adu] of user request $u$ \\
        $\dem{a, d}$ & aggregate demand [\adu] for $a$ on \dc $d$\\
        $\xi^a(f^a_i,d)$ & converts \dc demand [\adu] to capacity [\ecu]\\
        $\xi^a(e^a_{ij},e_{mn})$ & converts link demand [\adu] to bandwidth [\bwu]\\
        \midrule
        $\userv{i}{d}$ & amount [\adu] served on $d$ for $f^a_i$ of $u$ 
        & \multirow[b]{2}{*}{\llap{\color{blue}\raisebox{1ex}{\shortstack[r]{Decision\\Variables}}}} \\
        $\userp{ij}{st}$ & amount [\adu] served on $\pst$ for $e^a_{ij}$ of $u$ \\
        $\appv{i}{d}$ & size [\adu] of $f^a_i$ on $d$ \\
        $\appp{ij}{st}$ & size [\adu] of $e^a_{ij}$ on $\pst$ \\
        \bottomrule
    \end{tabularx}
\end{table}

\subsection{User Requests}\label{subsec:requests}
We denote by $\Set U$ the set of user requests and by $\Set A$ the set of applications. Each user request $u \in U$ is for an application $a(u) \in \Set A$, at a point of presence location, $d(u) \in \Set D$.
We denote by $\Set U(a)$ the set $\set{u \in \Set U \mid u(a)=a}$ and by $\Set U(a,d)$ the set $\set{u \in \Set U(a) \mid d(u)=d}$. The size of the user's request, denoted by $\dem{u}$, is specified in ``units of work'' that we call \emph{Application Demand Units (\Adu)}. The amount of traffic generated on the user's behalf, as well as the consumed resources are proportional to $\dem{u}$. We denote by $\dem{a, d}$ the aggregate demand for application $a$ at \dc $d$, namely, $\sum_{u \in \Set U(a,d)} \dem{u}$.

\subsection{Application Model}\label{subsec:applications}

\begin{figure}[h]
\includegraphics[width=\linewidth,page=31,trim=0cm 9.5cm 1cm 2.5cm,clip]{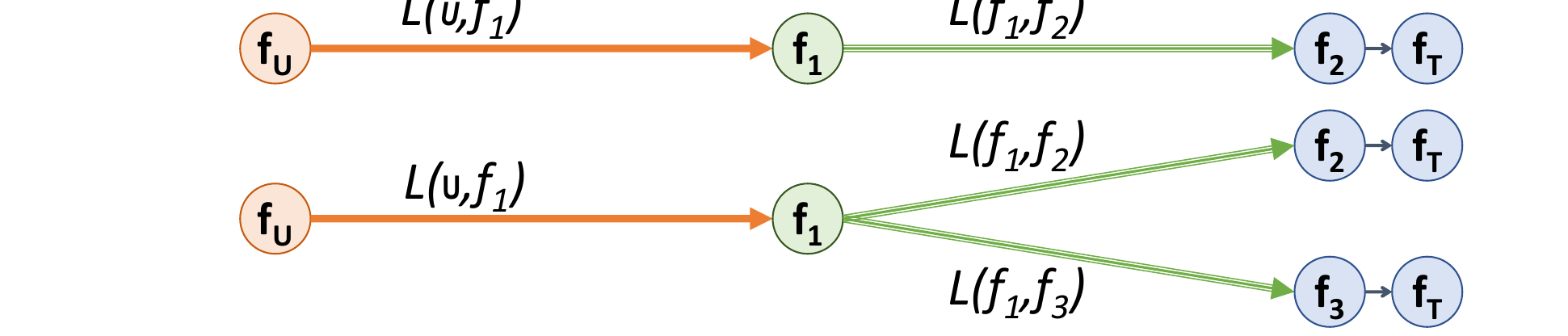}\caption{Applications topologies: The first node is always the UE, that should be placed at \ignore{. It's location is determined by} the user location. Similar to~\cite{MunkRostRackeSchmid-relax-2021,MaoSYY-INFOCOM2022}, latency constraints are specified between every pair of functions.}\label{fig:app-topology}
\end{figure}

Application SFC topologies are modeled as directed weighted graphs. The graph of an application $a \in \Set A$ is denoted by $G(\Set F^a, \Set E^a)$, where $\Set F^a$ is the set of functions in the application's SFC and $\Set E^a$ is its logical links. We consider two common types of application topologies: a \emph{chain} and a \emph{tree} as shown in \cref{fig:app-topology}\footnote{In future work, we plan to extend our results to general graphs by considering graph tree decomposition similarly to~\cite{RostDohneSchmid-parameterized-2019}.}. The root of each application topology is a fictitious function $f_{\ue}$ representing User Equipment (UE) of the user calling the application. A fictitious \emph{terminator} function $f_{\term}$ is added to each leaf of the topology.

We denote maximal tolerable latency constraint for each application link, denoted by $L(e^a_{ij}) = L(f^a_i, f^a_j)$. $L(f^a_{\ue}, f^a_i)$ represents the maximal latency allowed between the user and the first function $f^a_i$ of the application. The terminator has no latency constraints, that is $L(f^a_j, f^a_{\term}) = 0$ for all $f^a_{j \neq \term} \in \Set F^a$. 


\subsection{Problem Definition}\label{subsec:problem}\label{sec:milp}

\Cref{tab:1} describes the offline SFC Deployment Problem (\cref{prob:sfcd}). \Cref{eq:lp} offers a formal MILP. 

 %
{\floatstyle{boxed}\restylefloat{figure}
\begin{figure}
\Small\phantomsection\label[prob]{prob:sfcd}
\setlist[enumproblem]{label=(O\arabic*), align=left, leftmargin=1em, labelwidth=!}
\noindent\blue{\emph{Given the following inputs:}} 
\begin{enumproblem}[label=(I\arabic*)]
    \item \label{i1} Substrate network $G(\Set D, \Set E)$.
    \item \label{i2} Set of applications $\set{G(\Set F^a, \Set E^a)}_{a \in \Set A}$.
    \item \label{i3} Set of users $\Set U$ and their application requests.\footnote{All user requests are provided as input, as this is an \emph{offline} version of the problem.} 
\end{enumproblem}
\blue{\emph{Find the following outputs:}}
\begin{enumproblem}[label=(O\arabic*)]
    \item \label{o1} \textbf{Function placement.} Define where service functions are deployed and what sizes (flavors) are used. The variable $\appv{i}{d}$ denotes the size (in \adus) of $f^a_i$ on $d$.
    
    \item \label{o2} \textbf{Routing.} Select which substrate paths are used between the deployed functions. The variable $\appp{ij}{st}$ denotes the size (in \adus) of traffic on path $\pst$, between $f^a_i$ (on $s$) and $f^a_j$ (on $t$).
    
    \item \label{o3} \textbf{User allocation.} Allocate user requests onto the deployed functions. 
    The discrete variable $\userv{i}{d} \in \set{0, \dem{u}}$ denotes the amount of \adus served by the instance $f^a_i$ on \dc $d$ for user request $u$. 
    
    \item \label{o4} \textbf{Request steering.} Define the forwarding paths for the request traffic. The discrete variable $\userp{ij}{st} \in \set{0, \dem{u}}$ denotes the amount of \adus served by the substrate path $\pst$ to accommodate user's $u$ traffic between  $f^a_i$ and $f^a_j$.
\end{enumproblem}
\blue{\emph{Under the constraints:}}
\begin{enumproblem}[label=(C\arabic*)]
    \item \label{c1} \textbf{User location}. $f^{a(u)}_{\ue}$ must be located at $d(u)$, namely, $\userv{\ue}{d} = \dem{u}$ for $d=d(u)$ and 0 otherwise.
    
    \item \label{c2} \textbf{Latency}. Routing paths should meet the delay constraints for the traffic they carry. That is, if $\appp{ij}{st} > 0$ then $L(\pst) \le L(e^a_{ij})$.

    \item \label{c3} \textbf{Capacity}. The aggregate \ecus on each \dc and the aggregate \bwus on each substrate link must not exceed capacity. \ignore{\DL{Remove: } Note that $\xi()$ converts from \adus.}
\end{enumproblem}
\blue{\emph{With the following optimization goals (prioritized):}}
\begin{enumproblem}[label=(G\arabic*)]
    \item \label{g1} \textbf{Minimize rejection}. As many user requests as possible are deployed.
        
    \item \label{g2} \textbf{Minimize cost}. The overall cost of allocating the required resources on the substrate nodes and links is minimal. Note that minimizing rejections has priority over this goal, namely we are willing to incur a higher cost if more user requests are deployed.
\end{enumproblem}
\caption{SFC Deployment Problem (\cref{prob:sfcd})\label{tab:1}}
\end{figure}}



\Cref{lp0} states that each individual user request traffic must not split. \Cref{lp1} defines the relation between user allocation \cref{o3} and request steering \cref{o4} and \cref{lp2,lp3} aggregate these to define function placement \cref{o1} and routing \cref{o2}. The user location \cref{c1} ($f^a_{\ue}$ placement) and the latency \cref{c2} constraints are expressed in \cref{lp4,lp5}. In our model, the $L()$ expressions are constants. Therefore \cref{lp5} is linear similarly to~\cite{MaoSYY-INFOCOM2022, BehraveshHCR-TNSM2021,ChowdhuryRB-TON2012}.

\subsubsection*{Capacity constraints}
\begin{figure}
    \centering
    \includegraphics[width=0.7\linewidth,page=30,trim=3.9cm 11.5cm 3.9cm 1.8cm,clip]{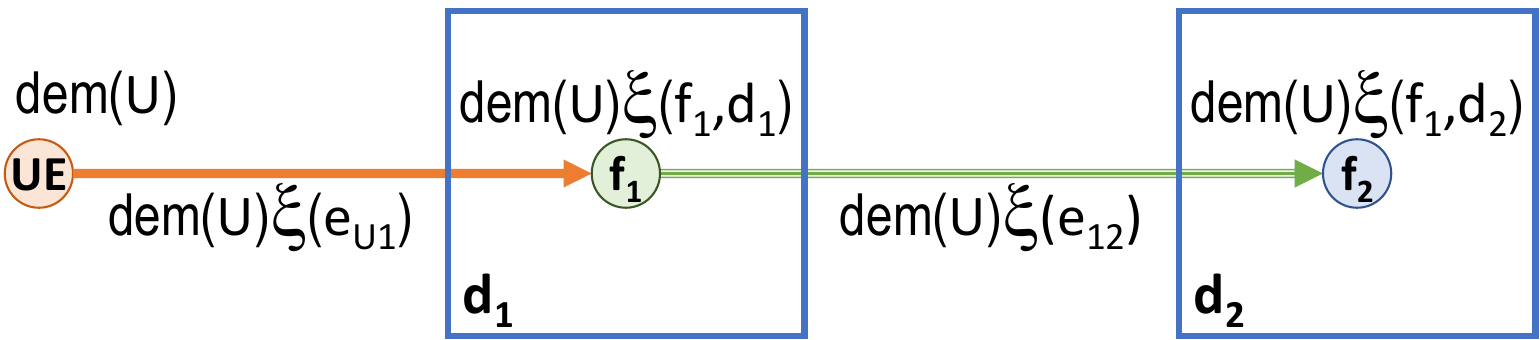}
    \caption{Converting demand to resources}\label{fig:mult}
\end{figure}

In order to express the capacity constraints \cref{c3}, we need to convert from a demand size to a resource requirement. The amount of resources required to serve one \adu depends on the substrate resources used to serve the application. We use a conversion function, denoted by $\xi()$, to convert \adus to \ecus and \bwus, as shown in  \cref{fig:mult}. 
The amount of resources consumed by a function $f^a_i$ that is hosted on \dc $d$, when serving one \adu of application $a$, is $\xi^a(f^a_i,d)$ \ecus. Similarly, the amount of resources consumed on a substrate link $(m,n)$ that is forwarding traffic of the logical link $e^a_{ij}$, when serving one \adu of application $a$, is $\xi^a(e^a_{ij}, e_{mn})$ \bwus. These conversion factors allow expressing variability in function sizes and inter-function traffic within a single application topology. They also allow expressing the efficiency of hosting a particular function on a particular \dc, e.g., if the \dc has specialized hardware suited to run the function. Finally, we define $\xi() = 0$ for logical links and nodes involving the fictitious functions $f_{\ue}$ or $f_{\term}$.   

\Cref{lp6,lp7,lp8} express the node and link capacity constraints. \Cref{lp7} defines a helper variable to compute load on a physical link due to the logical path traffic allocation. 

\begin{formulation}\caption{SFC Deployment Problem (MILP)}\label[milp]{eq:lp}
\Small
{\color{blue}\textbf{Given}}
$G(\Set D, \Set E), \set{G(\Set F^a, \Set E^a)}_{a \in \Set A}, \Set U, \xi()$,~
{\color{blue}\textbf{minimize}}
\addtolength{\belowdisplayskip}{-15pt}
\begin{multline}\label{eq:g2}
    \cost{\solution} = \cost{d} \cdot \smashoperator[l]{\sum_{a \in \Set A}}\smashoperator[r]{\sum_{f^a_i \in \Set F^a}} \xi^a(f^a_i,d) \appv{i}{d} \\
    + \cost{m,n} \cdot \smashoperator[l]{\sum_{a \in \Set A}} \smashoperator[r]{\sum_{e^a_{ij} \in \Set E^a}} \xi^a(e^a_{ij},e_{mn}) \appe{ij}{mn}{s}{t}    
\end{multline}
\addtolength{\belowdisplayskip}{15pt}
\blue{\textbf{such that}}
\begin{align}
\label{lp0}
\userv{i}{d}, \userp{ij}{st} &\in \set{0, \dem{u}}
\\
\label{lp1}
\userv{i}{s} &= \sum_{\mathclap{\pst \in \Pst}}\userp{ij}{st} = \userv{j}{t}
\\
\label{lp2}
\appv{i}{d} &= \smashoperator{\sum_{u \in \Set U(a)}}\userv{i}{d}
\\
\label{lp3}
\appp{ij}{st} &= \smashoperator{\sum_{u \in \Set U(a)}} \userp{ij}{st}
\\
\label{lp4}
\userv{\ue}{d(u)} &= \dem{u} 
\\
\label{lp5}
0 &\le \appp{ij}{st}(L(e^a_{ij})-L(\pst))
\\
\label{lp6}
\capacity{d} &\ge \smashoperator[l]{\sum_{a \in \Set A}}\smashoperator[r]{\sum_{f^a_i \in \Set F^a}} \xi^a(f^a_i,d) \appv{i}{d}
\\
\label{lp7}
\appe{ij}{mn}{s}{t} &= \sum_{\mathllap{\pst} \in \mathrlap{\Pst \text{ s.t. } e_{mn} \in \pst}} \appp{ij}{st}
\\
\label{lp8}
\capacity{e_{mn}} &\ge \smashoperator[l]{\sum_{a \in \Set A}} \smashoperator[r]{\sum_{e^a_{ij} \in \Set E^a}} \xi^a(e^a_{ij},e_{mn}) \appe{ij}{mn}{s}{t}
\end{align}
{\color{blue}$\forall u \in \Set U$, $\forall a \in \Set A$, $\forall d,s,t \in \Set D$, $\forall e_{mn} \in \Set E$, $\forall \pst \in \Pst$, $\forall f^a_i \in \Set F^a$, $\forall e^a_{ij} \in \Set E^a$}
\end{formulation}

\subsubsection*{Feasibility}
Let $\solution$ denote an assignment of $\appv{i}{d}$, $\appp{ij}{st}$, $\userv{i}{d}$, and $\userp{ij}{st}$ for all users. 
These variables correspond to the outputs \cref{o1,o2,o3,o4} of \cref{prob:sfcd}. Therefore $\solution$ defines a deployment that is a solution to \cref{prob:sfcd}. If $\solution$ satisfies \cref{eq:lp} then the deployment is \emph{feasible}, namely it satisfies \cref{c1,c2,c3}. 

\ignore{
\begin{define}[feasible deployment]
A deployment $\solution$ is \emph{feasible} if it satisfies \crefrange{lp0}{lp8}.
\end{define}
}
\subsubsection*{Optimization goal}
\cref{eq:lp} defines the goal of minimization as finding a feasible deployment with minimized total cost of hosting and traffic. This readily addresses \cref{g2}. To address the rejection minimization goal \cref{g1}, rather than adding new decision variables to indicate which users are rejected, we utilize the cost minimization optimization. We add a \emph{fictitious} substrate \dc, $\dfict$, to accommodate all rejected user's demands, but at a prohibitive cost. We set $\capacity{\dfict} \gets \infty$, $\xi(f, \dfict) \gets \max_{a,i,d} \xi^a(f^a_i, d)$, and $\cost{\dfict} \gets M \cdot \max_{d} \cost{d}$, where $M$ is a large constant.  We add substrate links from all \dcs to $\dfict$ and set their cost to zero and their capacity to $\infty$. Now, there is always a feasible solution to \cref{eq:lp}, as deploying all function and all user request to $\dfict$ is a feasible deployment. Moreover, it is more costly to deploy a user on $\dfict$ than any other alternative. Thus, the minimal cost, feasible allocation is the optimal solution to \cref{prob:sfcd}. 

\subsubsection*{Scalability}
A MILP solver (e.g., Gurobi \cite{gurobi}) can be used to find an exact optimal solution to \cref{prob:sfcd}, however, \cref{eq:lp} does not scale well. The number of variables is proportional to the number of users and to the number of substrate paths. Both these numbers can be high enough to render the solver approach impractical. In the next sections, we employ several heuristics that address the scalability issue.

\section{Our Solution}\label{sec:solution}

In this section we describe \cref{alg:prano}, our heuristic solution to \cref{prob:sfcd}. As shown in \cref{alg:prano}, it comprises two steps: (1) solving an LP relaxation~\cref{eq:flow} of MILP~\cref{eq:lp} and (2) greedy rounding of the fractional solution.\ignore{ to deploy \mbox{individual requests.}}

\ignore{We solve \cref{prob:sfcd} in two steps: first, as described in \cref{sec:flow}, we find the optimal fractional function placement and routing by \emph{relaxing} the MILP to an LP. Then, as described in \cref{sec:heur3}, we employ a greedy method to \emph{round} the solution and find the user allocation and request steering. These two steps are inspired by the online version of \cref{prob:sfcd}. 
In this section we provide our heuristic solution to \cref{prob:sfcd}, called \ref{alg:prano}. } \ignore{We solve \cref{prob:sfcd} in two steps: first, as described in \cref{sec:flow}, we find the optimal fractional function placement and routing by \emph{relaxing} the MILP to an LP. Then, as described in \cref{sec:heur3}, we employ a greedy method to \emph{round} the solution and find the user allocation and request steering. These two steps are inspired by the online version of \cref{prob:sfcd}. 
We significantly reduce the number of variables in the LP by (1) aggregating similar user requests and (2) employing a heuristic to remove path enumerations (see \cref{sec:path}). We prove a \emph{deterministic} bound on the number of users that might be rejected due to relaxation and rounding (see \cref{sec:analysis}).}

\begin{algorithm}[!h]
    \Small
    \floatname{algorithm}{Alg.}
    \renewcommand{\thealgorithm}{PRANOS}
    \caption{Practical Near-Optimal SFC Deployment}\label[pranos]{alg:prano}
    \begin{algorithmic}[1]
        \Ensure{Solution to \cref{prob:sfcd}}
        \Statex \vspace{0.2\baselineskip}\hrule
        
        \State Solve \cref{eq:flow}, find the optimal $\fraqopt$
        \Comment{\Shortstack{function placement and routing\\(see \Cref{sec:flow})}}
        \State Execute \cref{alg:heur3} with $\fraqopt$
        \Comment{\Shortstack{allocation and request steering\\(see \cref{sec:heur3})}}
    \end{algorithmic}
\end{algorithm}

\subsection{Function Placement and Routing}\label{sec:flow}

We transform \cref{eq:lp} MILP to \cref{eq:flow} LP. The transformation is based on the observation that \cref{eq:lp}
is akin to a multi-commodity flow problem~\cite{Karakostas-fractionalMCF2008} with link-level latency restrictions. 
There is a commodity for each combination of source \dc $s$, logical link $e^a_{ij}$, and substrate link $e_{mn}$. We express placement and routing using two types of flow-based decision variables, \emph{direct} flow, denoted by $\dflow{s}{ij}{mn}$, and \emph{transient} flow, denoted by $\tflow{s}{ij}{mn}$. 
The flows are formally defined in \cref{def:dflow,def:tflow}. 

\begin{table}[b]
    \caption{Notations}
    \label{tab:notations2}
    \centering\Small
    \begin{tabularx}{\columnwidth}{@{}lXr@{}}
        \toprule
        \textbf{Notation} & \textbf{Description} \\ 
        \midrule 
        $\delta{st}$ & Shortest path latency between $s$ and $t$ 
        & \multirow[b]{2}{*}{\llap{\color{blue}\shortstack[r]{Substrate\\Network}}} 
        \\
        $\preasonable{s}{mn}$ & $\ge 0$ if $e_{mn}$ allowed in $\Pst$ \\
        $L(\pst)$ & latency of path $\pst$ \\
        $L(\Pst)$ & maximal latency of \emph{any} path $\pst \in \Pst$ \\
        \midrule 
        $\appv{j}{d}$ & size [\adu] of $f^a_j$ on $d$ 
        & \multirow[b]{2}{*}{\llap{\color{blue}\raisebox{1ex}{\shortstack[r]{Decision\\Variables}}}}
        \\
        $\appbw{ij}{mn}$ & size [\adu] of $e^a_{ij}$ on $e_{mn}$ \\
        \raisebox{-1ex}{$\dflow{s}{ij}{m,n}$} & direct traffic embedding of application link $e^a_{ij}$ on substrate link  $e_{mn}$ from \dc $s$ \\
        \raisebox{-1ex}{$\tflow{s}{ij}{m,n}$} & transient traffic embedding of application link $e^a_{ij}$ on substrate link $e_{mn}$ from \dc $s$ \\
        \bottomrule
    \end{tabularx}
\end{table}

\begin{define}\label{def:dflow}
    $\dflow{s}{ij}{mn}$ is the amount of traffic (in \adu) from function $f^a_i$ towards function $f^a_j$ on \dc $n$ over the substrate link $e_{mn}$ for the traffic that originates on a source \dc $s$. We term such traffic \textbf{direct} $s$-flow to \dc $n$.
\end{define}
\begin{define}\label{def:tflow}
    $\tflow{s}{ij}{mn}$ is the amount of traffic (in \adu) from function $f^a_i$ towards function $f^a_j$ on some \emph{other} \dc $t$, where $n \neq t \neq m$ over the substrate link $e_{mn}$ for the traffic that originates on a source \dc $s$. We term such traffic \emph{transient} $s$-flow through \dc $n$. 
\end{define}

Embedding of an SFC through is done through a combination of direct and transient flows. We allow flows to split along the path towards different $f^a_j$ locations (and even merge back later on the path). 
Note that, while fractional flows define function placement and routing; individual request steering should be unsplittable. We address this in the second, rounding stage, described in \cref{sec:heur3}.

\ignore{
\Cref{fig:psd} shows the two types of flows -- direct flows as solid lines and transient flows as dashed lines. Both are sourced from the network function $f^a_i$ that is located on \dc $s$. The difference between them is that $e_{mn}$ is the \emph{last} link for the direct flow and an \emph{intermediate} link for the transient 
flow. The flow along $\pst[sd]$ (which is transient for all but the last hop)  is given by 
$
\set{\tflow{s}{ij}{sn_1}, \tflow{s}{ij}{n_1n_2}, \tflow{s}{ij}{n_2n_3}, \ldots, \dflow{s}{ij}{n_kd}}.
$ 

\Cref{fig:def1} shows an embedding of an application chain through a combination of direct and transient flows. We allow flows to split along the path towards different $f^a_j$ locations (and even merge back later on the path). Note that, while fractional flows defines function placement and routing; individual request steering should be unsplittable. We address this in the second, rounding stage, described in \cref{sec:heur3}.}

A flow on $e^a_{ij}$ can be embedded entirely in $s$, namely $f^a_i$ and $f^a_j$ may be collocated at $s$. 
In this intra-\dc case, since the application link $e^a_{ij}$ is embedded within \dc $s$, we assume that the bandwidth and latency constraints for $e^a_{ij}$ are met. 
On the other hand, flow loops are not allowed, namely, traffic coming from $f^a_i$ on $s$ cannot leave $s$ and go back to it.

\begin{formulation}\caption{Function Placement and Routing (LP)}\label[flow]{eq:flow}
\Small
\blue{\textbf{Given}} $G(\Set D, \Set E), 
\set{G(\Set F^a, \Set E^a)}, 
\set{\dem{d,a}}, \xi()$,~
\blue{\textbf{minimize}}
\addtolength{\belowdisplayskip}{-5pt}
\begin{multline}\label{eq:cost}
    \cost{\chi} = 
    \sum_{d \in D}\cost{d}\sum_{a \in \Set A} \sum_{f^a_j \in \Set F^a} \xi^a(f^a_j,d) R_d(f^a_j)
    \\ +
    \sum_{m,n \in \Set D} \cost{m,n} \sum_{a \in \Set A} \sum_{e^a_{ij} \in \Set E^a} \xi^a(e^a_{ij},e_{mn}) R_{mn}(e^a_{ij})
\end{multline}
\addtolength{\belowdisplayskip}{5pt}
\blue{\textbf{such that}}
\begin{align}
\label{dflow} 
\sum_{s,m \in \Set D} \dflow{s}{ij}{m,d} &= \sum_{n \in \Set D} \left( \dflow{d}{j,k}{d,n} + \tflow{d}{j,k}{d,n} \right)
&\qquad\qquad\quad
\mathllap{\color{blue}\substack{\forall e^a_{ij},e^a_{jk} \in \Set E^a}} 
\\
\label{tflow} 
\sum_{m \in \Set D} \tflow{s}{ij}{m,d} &= \sum_{n \in \Set D}  \left( \dflow{s}{ij}{d,n} + \tflow{s}{ij}{d,n} \right)
&\mathllap{\color{blue}\substack{\forall e^a_{ij},e^a_{jk} \in \Set E^a}} 
\\
\label{loop} 
\dflow{s}{ij}{m,m} &= \dflow{s}{ij}{m,s} = \tflow{s}{ij}{m,s} = 0 
&\mathllap{\color{blue}\substack{\forall m \neq s \in \Set D}}
\\
\label{noselftran} 
\tflow{s}{ij}{m,m} &= \tflow{s}{ij}{s,s} = 0
&\mathllap{\color{blue}\substack{\forall s \neq m \in \Set D}} 
\\
\label{term_col} 
\dflow{s}{i,\term}{mn} &= 0
&\mathllap{\color{blue}\substack{\forall m \neq n \in \Set D}}
\\
\label{demand_sat} 
\dem{d,a} &= \sum_{n \in \Set D} \left( \dflow{d}{\ue,k}{d,n} + \tflow{d}{\ue,k}{d,n} \right) 
\\
\label{lat_reas1} 
0 &\le \dflow{s}{ij}{mn} \preasonable[]{s}{m,n}
\\
\label{lat_reas2} 
0 &\le \tflow{s}{ij}{mn} \preasonable[]{s}{m,n}
\\
\label{lat_path} 
0 &\le \dflow{s}{ij}{n,t} \left( L(e^a_{ij}) - L(\Pst) \right)
\\
\label{appfuncBW1} 
\appv{j}{d}
&= \sum_{e^a_{ij} \in \Set E^a}\sum_{s,m \in D} \dflow{s}{ij}{m,d} 
\\
\label{dcBW} 
\capacity{d} &\ge \sum_{a \in \Set A} \sum_{f^a_j \in \Set F^a} \xi^a(f^a_j,d) \appv{j}{d}
\\
\label{applinkBW} 
\appbw{ij}{mn}
&= \sum_{s \in D} \left( \dflow{s}{ij}{mn} + \tflow{s}{ij}{mn} \right) 
\\
\label{linkBW} 
\capacity{e_{mn}} &\ge \sum_{a \in \Set A} \sum_{e^a_{ij} \in \Set E^a} \xi^a(e^a_{ij},e_{mn}) \appbw{ij}{mn}
\end{align}
{\color{blue}$\forall a \in \Set A$, $\forall d,s,t,m,n \in \Set D$, $\forall e_{mn} \in \Set E$, $\forall \pst \in \Pst$, $\forall e^a_{ij} \in \Set E^a$}
\end{formulation}

\subsubsection*{Latency}\label{sec:lat}
To avoid path enumeration, we restrict every path $\pst \in \Pst$ to a subset of substrate links, $\set{e_{mn} \in \Set E \mid \preasonable[]{s}{m,n} \ge 0}$, where $\preasonable[]{s}{m,n}$ is set at the pre-processing time to values that guarantee $L(\Pst) \le \alpha \delta_{st}$, where $L(\Pst) = \max_{\pst \in \Pst} L(\pst)$, $\alpha$ is a constant, and $\delta_{st}$ denotes the shortest path latency between $s$ and $t$. This set does not depend on the destination $t$ (see \cref{sec:path} for details).
\subsubsection*{LP}\label{sec:lpflow}
\Cref{eq:flow} is our relaxed formulation for the function placement and routing of \cref{prob:sfcd}. 
\Cref{dflow,tflow} define flow preservation, replacing \cref{lp1}. For a logical link $e^a_{ij}$, the flow $f^a_i \to f^a_j$ into a node, both direct \cref{dflow} and transient \cref{tflow}, must continue as either direct flow or transient flow out of the node.\ignore{(see \cref{fig:dflow_preserv,fig:tflow_preserv}).} The difference is that with direct flow the node must be hosting the function $f^a_{j}$ and the outgoing traffic is on a different logical link. \Cref{loop,noselftran,term_col} ensure that there are no flow loops, transient traffic is only \emph{inter}-\dc, and that the logical link to the fictitious terminator $f^a_{\term}$ is always \emph{intra}-\dc. \Cref{demand_sat} requires all demand to be served; it is similar to \cref{lp4}, but uses the \emph{aggregated} demand for $a$ on $d$. 

\Cref{lat_path,lat_reas1,lat_reas2} define latency constraints, replacing \cref{lp5}. \Cref{lat_reas1,lat_reas2} enforce the link restrictions, ensuring that flows on restricted links is $0$. Any $e^a_{ij}$-flow over $\pst \in \Pst$ ends with a  direct flow; therefore, \cref{lat_path} enforces the path latency constraints. 
\Cref{appfuncBW1} states that the size of $f^a_j$ on $d$, $\appv{j}{d}$, is given by the aggregate direct $e^a_{ij}$-flow into a node $d$. \Cref{applinkBW}, replacing \cref{lp7}, is the aggregate link flow, denoted by $\appbw{ij}{mn}$; it is independent of $t$ or $\pst$. The corresponding \dc and link capacity constraints are stated in \cref{dcBW,linkBW}; similar to \cref{lp6,lp8}, $\xi()$ is used to convert \adus to \ecus and \bwu. 

\subsubsection*{Feasibility}\label{sec:feas}
Let $\fraqopt$ denote an assignment of $\dflow{s}{ij}{m,d}$ and $\tflow{s}{ij}{m,d}$ for all applications. $\appv{i}{d}$ provides function placement and sizes, and the flows of $\fraqopt$ define routing. 
We say that $\fraqopt$ that satisfies \crefrange{dflow}{linkBW} of \cref{eq:flow} is \emph{fractional feasible}.
\ignore{
\begin{define}[feasible flow]
A flow $\fraqopt$ is \emph{feasible} if it satisfies \crefrange{dflow}{linkBW}.
\end{define}}

The cost of an allocation $\fraqopt$ \cref{eq:cost} is given by multiplying the total capacity requirements \cref{dcBW,linkBW} by the cost of processing and bandwidth. 
The minimal cost fractional feasible flow provides optimal placement and routing solution to a relaxed \cref{prob:sfcd}. It is relaxed in the sense that $\appv{i}{d}$ is not required to be integer and can be smaller than $1$. Note that minimal cost feasible $\fraqopt$ is a \textit{fractionally optimal} solution for MILP \cref{eq:lp}, because any feasible deployment $\solution$ can be transformed by aggregation to a feasible flow $\fraqopt$ and $\cost{\chi} \le \cost{\solution}$ when total cost of \textit{both} deployed and denied requests is considered \footnote{If cost of deployed requests \textit{only} is considered, then $\cost{\chi} \ge \cost{\solution}$}.

\ignore{\Cref{th:fraqopt} state that $\fraqopt$ is fractionally \emph{optimal}.
\begin{theorem}\label{th:fraqopt}
    For an optimal solution $\solution$ to MILP \cref{eq:lp} and fractionally optimal solution $\fraqopt$ to LP \cref{eq:flow}, $\cost{\chi} \le \cost{\solution}$ and the amount of denied demand (i.e., requests allocated at $\dfict$) by $\fraqopt$ is not greater than the amount of demand denied by $\solution$.
\end{theorem}
\begin{IEEEproof}
    Any feasible deployment $\solution$ can be transformed by aggregation to a feasible flow $\fraqopt$. The results follow. 
\end{IEEEproof}}

        

\subsection{User Allocation and Request Steering}\label{sec:heur3}
In this subsection, we describe how to round fractionally optimal solution $\fraqopt$ found by solving LP \cref{eq:flow}, to obtain a feasible unsplittable solution $\solution$ to \cref{prob:sfcd}. 

We employ a greedy approach, as shown in \cref{alg:heur3}, to embed one user request at a time. In each greedy step we search for a single user embedding that complies with $\fraqopt$. Note that this is different from the greedy heuristic solutions typically applied to the VNF embedding problem~\cite{SCHARDONG2021107726}. Although seemingly similar, the latter locally optimizes for a given user request, while we use $\fraqopt$ as a ``plan'' that is already globally optimized. 

\newcommand{\embapp}{Embed-App-From-DC}
\newcommand{\emblink}{Embed-Link-From-DC}

\begin{algorithm}
\Small
\caption{User Allocation and Request Steering}\label{alg:heur3}
\begin{algorithmic}[1]
    \Require{$G, \chi, \Set U$}
    \Ensure{Per user \bwu and \ecu allocation, set of rejected users}
    
    \Statex \vspace{0.5\baselineskip}\hrule
    
    \State $\partreject \gets \emptyset$ \Comment{rejected users due to rounding}
    \For{$u \in \Set U$}\label{lin:mainloop3}
        \State $\set{\pst[ij]}, \set{v(f^a_i)}, \lambda \gets$\Call{\embapp}{$a(u),d(u)$}\label{lin:trytree3}
        
        \If{$\lambda \ge \dem{u}, u$} \Comment{sufficient resources}
            \State \Call{Allocate}{$u, \set{\pst[ij]}, \set{v(f^a_i)}, \dem{u}$}\label{lin:allocate3}
        \Else\label{lin:fail3} \Comment{insufficient resources}
            \State \Call{Allocate}{$u, \set{\pst[ij]}, \set{v(f^a_i)}, \lambda, u$}\label{lin:allocate4}
            \State $\partreject \gets \partreject \cup \set{u}$\label{lin:fail4}
            \EndIf
    \EndFor
    \State \Return $\partreject$

    \Statex \vspace{0.5\baselineskip}\hrule
    \Function{\embapp}{$a,s$}
        \Statex\emph{Return paths, user function allocation, and maximal embedded \adus.}
        
        \State $v(f^a_{\ue}) \gets s$ \Comment{$f^a_{\ue}$ is embedded at $s$} 
        \For{$e^a_{ij} \in \Set E^a$ in BFS order}
            \State $\pst[ij], v(f^a_j), \lambda_{ij} \gets$\Call{\emblink}{$e^a_{ij}, v(f^a_i)$}\label{lin:trylink3}
        \EndFor
        \State \Return $\set{\pst[ij]}, \set{v(f^a_i)}, \min\set{\lambda_{ij}}$
    \EndFunction
    
    \Statex \vspace{0.5\baselineskip}\hrule
    \Function{\emblink}{$e^a_{ij}, s$}
        \Statex \emph{Return path for $e^a_{ij}$, \dc for $f^a_j$, and maximal embeddable \adus.}
        \State $\pst[] \gets \emptyset, m \gets s, \lambda \gets \infty$
        \While{$\exists n \in \Set D$, s.t., $\tflow{s}{ij}{mn} > 0$}\label{lin:tpath1}
            \Comment{transient flow}
            \State $\pst[] \gets \pst[] \cup \set{(m,n)}$
            \State $\lambda \gets \min\set{\lambda, \tflow{s}{ij}{mn}}$\label{lin:tlambda}
            \State $m \gets n$
        \EndWhile\label{lin:tpath2}
        \State Choose $n \in \Set D$, s.t., $\dflow{s}{ij}{mn} > 0$\label{lin:dpath1} \Comment{direct flow}
        
        \State \Return $\pst[] \cup \set{(m,n)}, n, \min\set{\lambda, \dflow{s}{ij}{mn}}$\label{lin:dlambda} 
    \EndFunction
            
    \Statex \vspace{0.5\baselineskip}\hrule
    \Function{Allocate}{$u, \set{\pst[ij]}, \set{v(f^a_i)}, \lambda, u$}
        \Statex\emph{Allocate $\lambda$ \adus to $u$.}
        
        \For{$\pst[ij] \in \set{\pst[ij]}$}
            \If{$\lambda \ge \dem{u}$} 
                \State steer $u$'s $e^a_{ij}$ traffic over $\pst[ij]$, reserve $\lambda \cdot \xi^a(e^a_{ij},e_{mn})$ \bwus
            \EndIf
            \For {$e_{mn} \in \pst[ij]$}
                \If{$n = v(f^a_j)$}\State 
                    $\dflow{s}{ij}{mn} \gets \dflow{s}{ij}{mn} - \lambda$
                \Else\State 
                    $\tflow{s}{ij}{mn} \gets \tflow{s}{ij}{mn} - \lambda$
                \EndIf
            \EndFor
        \EndFor
        \If{$\lambda \ge \dem{u}$} \For{$f^a_i \in \Set F^a$}
            \State Allocate $\lambda \cdot \xi^a(f^a_i, v(f^a_i))$ \ecus for $u$ on $v(f^a_i)$
        \EndFor\EndIf
    \EndFunction
    \end{algorithmic}
\end{algorithm}

The main loop (\cref{lin:mainloop3}) goes over all users. For each user $u$, it calls Function~\textproc{Embed-Application-From-Node} to find an unsplittable allocation of $a(u)$ (steering paths and function locations) rooted at the user location $d(u)$ (\cref{lin:trytree3}). 

Function~\textproc{\embapp} embeds the application functions onto substrate \dcs and calls Function~\textproc{\emblink} to embed each logical link $e^a_{ij}$ onto a substrate path $\pst[ij]$. The functions and links are embedded in BFS order, starting from the user location, to ensure each embedded path continues from the end of the previous one. 

Function~\textproc{\emblink} implements a DFS search to greedily find a path. Since $\fraqopt$ is a min-cost flow, there are no flow loops; moreover, as long as $\fraqopt$ is not zero, a path exists. Note that due to the LP relaxation, the residual flow on the path may be smaller than the user's demand. Every path starts with zero or more transient flow links (\crefrange{lin:tpath1}{lin:tpath2}) and ends with a single direct flow link (\cref{lin:dpath1}). The function returns the path $\pst[ij]$, its destination \dc (location of $F^a_j$) and the maximal \adus that the path can support. 

When the embedding is complete, Function~\textproc{\embapp} computes the available \adus for the entire application. The main loop then calls Function~\textproc{Allocate} to update $\fraqopt$ and make the allocation. If the available \adus are sufficient to support the entire user demand (\cref{lin:allocate3}) then the functions of $a(u)$ are allocated, its steering is set, and $\fraqopt$ is updated to the remaining residual flow. If the available \adus are insufficient (\cref{lin:allocate4}), $u$ is rejected (\cref{lin:fail4}). In this case, no allocation or steering is made, however, Function~\textproc{Allocate} is still called (\cref{lin:fail4}) to update $\fraqopt$. This is to ensure that subsequent user request will not attempt to use the same embedding. 

\subsection{Analysis}\label{sec:analysis}

We observe that the number of decision variables in \cref{eq:flow} is
$2\abs{\Set V}\abs{\Set E}\sum_{a \in \Set A}\abs{\Set E^a})$, which is much smaller than that in \cref{eq:lp}.
It depends neither on the number of user requests, $\abs{\Set U}$, nor on the number of enumerated paths $\sum\abs{\Pst}$. Both demand and routing are now in terms of aggregate user flows and . our evaluation (\cref{sec:evaluation}) shows that we can handle millions of user requests on large topologies. 

\begin{theorem}\label{th:denied3}
\cref{alg:heur3} runs in $O(\abs{\Set V}\abs{\Set E^a})$ time per user $u$ and, at its conclusion, $\abs{\partreject} \le 2 \sum_{a \in \Set A}\abs{\Set V}\abs{\Set E}\abs{\Set E^a}$
\end{theorem}

\begin{IEEEproof}
    Function~\textproc{\emblink} finds loop-free paths, i.e., with length no larger than $\abs{\Set V}$. Unlike DFS, it never needs to back track, so the first non-zero link can be used for each next hop, hence it runs in $O(\abs{\Set V})$ time. It is called $O(\abs{\Set E^a})$ times for each user. 
    
    A user $u$ is added to $\partreject$ only if the $\lambda$ returned by \textproc{Embed-Application-From-Node} is less than $\dem{u}$ (\cref{lin:fail3}). Function~\textproc{\emblink} always sets $\lambda$ to the value of either a transient flow (\cref{lin:tlambda}) or a direct flow (\cref{lin:dlambda}) that is used for the embedding. Since $\lambda < \dem{u}$, \textproc{Allocate} would be called to allocate $\lambda$ resources (\cref{lin:allocate4}) and at least one of the variables in $\fraqopt$ will be zeroed by this allocation. Thus, each time a user demand is added to $\partreject$ at least one variable in $\fraqopt$ is zeroed. This implies that $\abs{\partreject}$ is less than the number of variables in $\fraqopt$ and the result follows. 
\end{IEEEproof}

In theory, the gap between splittable and unsplittable flow can be wide, however this gap assumes flows with arbitrary sizes. This gap is much smaller when the link and \dc capacities are far larger than the demand of a single user. Moreover, the deterministic bound provided by \cref{th:denied3} is independent of the number of users, therefore the fraction of rejected users diminishes as the number of users increases. Our evaluation (\cref{sec:evaluation}) shows very few users are rejected due to rounding. 

\subsection{Avoiding Path Enumeration}\label{sec:path}

Path enumeration requires that for every pair of substrate nodes $s,t$ we define a set of substrate paths $\Pst$. The number of variables in \cref{eq:lp} is proportional to the number of enumerated paths, $\sum\abs{\Pst}$. In general, $\abs{\Pst}$ grows exponentially with the size of the substrate network, even if we restrict $\Pst$ to include only shortest paths. Thus, path enumeration does not scale.

We propose a simple heuristic to overcome this problem. Our heuristic defines $\Pst$ only in terms of the links it uses. This limits the choice of paths in $\Pst$, but, as shown below, is still flexible enough to describe practical choices, for example, selecting all shortest $st$-paths.  

The constant $\preasonable{s}{m,n}$ denote whether the substrate link $e_{mn}$ can be used for \emph{any} path from $s$ to $t$. We use $\preasonable{s}{m,n} \ge 0$ to indicate that it can be used and $\preasonable{s}{m,n} < 0$ that it cannot. We now define $\Preasonable{s}$ to include \emph{all} possible substrate paths between $s$ and $t$ that use only links for which $\preasonable{s}{m,n} \ge 0$. Note that $\preasonable{s}{m,n}$ can be computed at a pre-processing time and does not depend on the decision variables of \cref{eq:lp}. It does not need to be expressed as a linear computation and can even be preset to express arbitrary path constraints. It can readily be extended to apply only to specific applications or even to specific application links. 


\subsubsection*{Shortest Path Only}

Recall that $\delta(s,t)$ denote the shortest path latency between \dcs $s$ and $t$. To include only shortest paths in $\Preasonable{s}$, we define: 
\begin{equation}\label{eq:sponly}
\preasonable{s}{m,n} =  \delta(s,n) - \left(\delta(s,m) + L(m,n)\right)
\end{equation}

Shortest path relaxation implies $\preasonable{s}{m,n} \le 0$ and $\preasonable{s}{m,n}=0$ only if $e_{mn}$ is on the shortest $s$ to $n$ path. Note that $t$ is not needed, that is, $\preasonable{s}{m,n} = \preasonable[]{s}{m,n} \quad\forall t \in \Set D$.

\subsubsection*{Geographic ``cabdriver'' paths} 
Assume all \dcs $d \in \Set D$ have a known geographic location $(d_x, d_y)$ and that $\delta_{st} \le \alpha\norm{st}$, where $\norm{st}$ is the geographic distance between $s$ and $t$. We define $\Preasonable{s}$ to include only shortest ``cabdriver'' paths on the geographic grid:
\begin{equation}\label{eq:geolatency}
    \preasonable{s}{m,n} = 
    \begin{cases}
        -1 & \text{if\phantom{or}} \abs{m_x - s_x} > \abs{n_x - s_x} \\&\text{or\phantom{if}} \abs{m_y - s_y} > \abs{n_y - s_y} \\
        \phantom{-}0 & \text{if\phantom{or}} \delta(s,m) + L(m,n) = \delta(s,n) \\
        \phantom{-}1 & \text{otherwise}
    \end{cases}
\end{equation}
In other words, we disallow the use of $(m,n)$ if it heads back towards \dc $s$. The $0$ case (overrides $-1$) is to explicitly allow using the link if it is on the shortest path from $s$ to $n$. Note that, again, $t$ is not needed, that is, $\preasonable{s}{m,n} = \preasonable[]{s}{m,n} \quad\forall t \in \Set D$.

\begin{figure}[h!]
    \centering
    \includegraphics[page=11,width=.45\columnwidth,trim=8.5cm 1.5cm 9cm 3cm,clip] {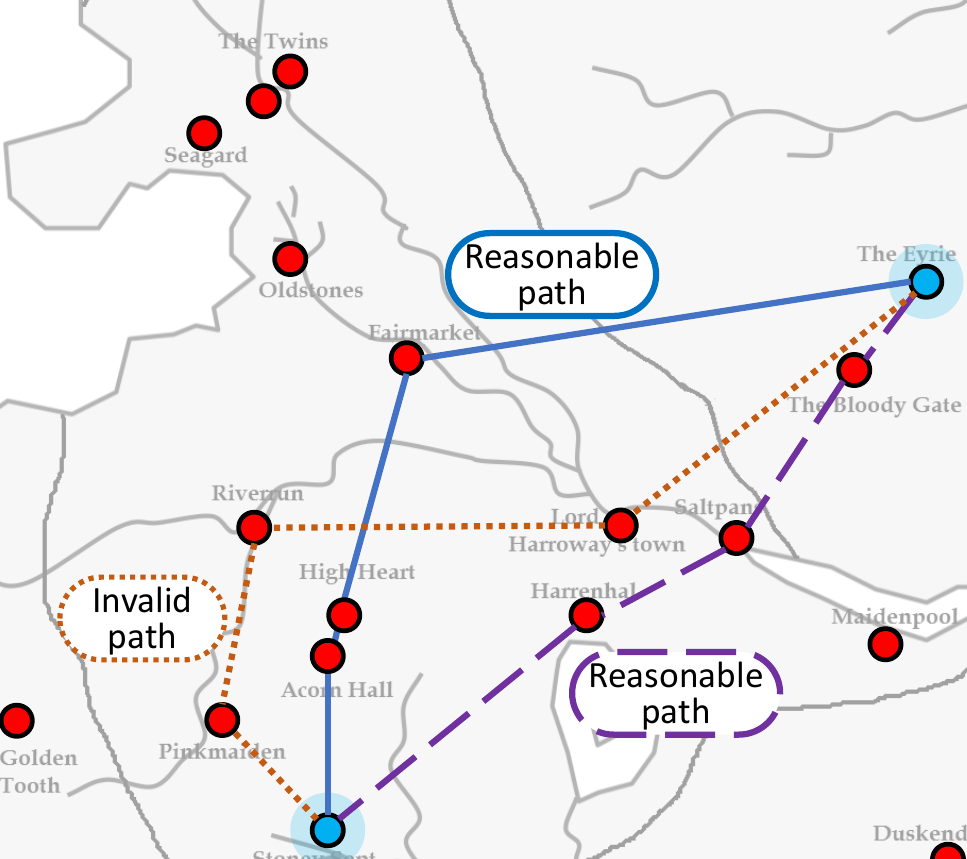}
    \caption{Geographically ``cabdriver'' paths. \Small Paths that head-back west towards a destination on the East are invalid.}
    \label{fig:geodistance}
\end{figure}

\subsubsection*{Path latency} 

Using \cref{eq:sponly}, we get $L(\Pst) = \delta(s,t)$. Using \cref{eq:geolatency} we get $L(\Pst) \le \alpha' \cdot\delta(s,t)$, for some conversion factor $\alpha'$. In our evaluation with use \cref{eq:geolatency}. 

\section{Evaluation}\label{sec:evaluation}

        


\begin{table}
    \caption{Details of the topologies}
    \label{tab:topologies}
    \centering\Small
    \begin{tabularx}{\columnwidth}{@{}lccX@{}}
        \toprule
        \makebox[2.2cm][l]{\textbf{Topology}}& \textbf{\# Nodes} & \textbf{\# Links}  & \textbf{Description} \\ 
        \midrule 
        40N60E & 40 & 60  & Random graph \cite{XiangEMN-2023} 
        \\
        100N150E& 100 & 150  & Random graph \cite{XiangEMN-2023} \\
        Citta Studi & 30 & 35 & Edge network topology \cite{XiangEMN-2023}\\
        5GEN & 78 & 100 & Realistic 5G/6G topology \cite{5GEN}\\
        \midrule 
    \end{tabularx}
    \begin{tabularx}{\columnwidth}{@{}Xccc@{}}
        \textbf{Parameter} & \textbf{Edge} &\textbf{Transport} (5GEN Only) & \textbf{Core} \\ 
        \midrule 
        Node Cap [\ecus] &  200K & 800K & 2.5M
        \\
        Node Cost (per \ecu) & 50 & 10 & 1 \\
        Link Cap [\bwus ] & 200K & 200K & 200K \\
        Link Cost (per \ecu) & 1 & 1 & 1 \\
        \bottomrule
    \end{tabularx}
\end{table}

In this section, we extensively evaluate~\ref{alg:prano} by comparing it to a state-of-the-art greedy heuristic HEU\_Cost~\cite{harutyunyan2019latency, harutyunyan2020latency} and the fractional solution $\fraqopt$ obtained by solving \cref{eq:flow}, which is also a theoretical lower bound for \cref{prob:sfcd}.

\subsubsection*{Network Substrate}
We use three different topology types: (1) a realistic 5G/6G topology reflective of 5G deployment in Madrid, Spain, generated by 5GEN~\cite{5GEN}, (2) a realistic mobile edge network topology in “Citt{\'a} Studi” area around Politecnico di Milano reported in~\cite{XiangEMN-2023}, and connected Erd\H{o}s-R{\'e}nyi random graphs~\cite{Erdos:1959:pmd} of the same size generated by the same method as the one reported in~\cite{XiangEMN-2023}.
\Cref{tab:topologies} summarizes physical substrate topologies that we used in our experiments.

\subsubsection*{Applications topology}
Inspired by~\cite{Sharma2021}, we explore applications (i.e., SFCs) with a relatively small number of VNFs, because these are the cases important in practice. In our experiments, the average number of VNFs per application instance is $4$.  We consider both chain and tree application topologies of variable sizes having different latency constraints, capacity requirements, and multipliers.\footnote{Note that decision variables grow linearly with the number of links in the topology. Thus, it has only a moderate impact on the algorithm performance.}

\subsubsection*{Users}
We evaluate~\ref{alg:prano}, our approach with a very large number of users ranging from $5K$ to $100K$ and to $1M$ users. We have two sets of experiments. In one set, we distribute user requests (i.e., SFC deployment requests) uniformly across the point of presence \dcs. In another set of experiments, we distribute user requests using truncated Zipf distribution, so that a few points of presence \dcs are extremely popular, which simulates ``hotspots'' that might correspond to large-scale events, daily patterns in traffic or failures in the physical network that overload some point of presence \dcs, while other are relatively under-loaded.

\subsubsection*{Latency Constraints}
We perform a comprehensive study on how application latency constraints impact~\ref{alg:prano}, HEU\_Cost~\cite{harutyunyan2020latency}, and fractionally optimal solution $\fraqopt$. We use a mix of two applications: \emph{Relaxed}, no latency restrictions on either application; \emph{Strict}, strict latency requirements on both applications that force VNFs to be placed either in the same \dc or in the same topology level (edge, aggregation, or core), but not across levels; \emph{Mixed}, one application is under strict latency constraints and the other one has no latency constraints. The mix of the applications is created randomly.

\subsubsection*{Execution Environment}
The simulations are implemented in Python with Gurobi~\cite{gurobi} mathematical optimization as the back-end solver.\footnote{Upon publication of our work,
we intend to release our implementation as open-source code to benefit the research community.} It is run on an Apple M1 8 x Cores CPU @ 3.2GHz with 16 GB RAM.

\subsubsection*{Experiments Structure}
Each experiment is defined by a triple: \textless \textit{Topology}, \textit{User Distribution}, \textit{Latency Constraints}\textgreater. The results given are averaged over $20$ experiments with a standard deviation of the number of rejected embedding requests, allocated demand, and execution times. Due to the lack of space, we cannot present this study in full. Rather, we show a subset of cases, sufficient to gain insights.  

\subsection{Results}\label{subsec:results}
In this subsection, we describe our results.
\subsubsection*{User Requests Rejection}

Figure~\ref{fig:40}(a), Figure~\ref{fig:100}(a), Figure~\ref{fig:citta}(a), Figure~\ref{fig:5gen}(a) show that for the hotspot \dc scenario modeled via allocating user requests to the point of presence \dcs using truncated Zipf distribution with parameter $a=1.2$ and relaxed latency restrictions which allow placement of neighboring functions in SFC anywhere in the physical network substrate, the number of user requests rejected by\mbox{~\ref{alg:prano}} is very close to the lower bound $\fraqopt$ and much smaller than that of HEU\_Cost. In this set of experiments the same SFC topology was requested by all users.

Figure~\ref{fig:40}(b), Figure~\ref{fig:100}(b), Figure~\ref{fig:citta}(b), Figure~\ref{fig:5gen}(b) show that for the same hotspot scenario as above, but when latency restrictions are strict, which requires to place neighboring functions in the same layer of the topology (e.g., edge, transport or core), there are significantly fewer options that global optimization of \ref{alg:prano} can exploit. Therefore all methods reject a considerably larger number of user requests and the gap between the different methods is much less pronounced. One can also notice that for the same load applied, for larger physical network topologies, a saturation point leading to rejecting user requests happens for a larger number of user requests, as one would expect.  

\begin{figure}
    \centering
    \includegraphics[width=.8\linewidth]{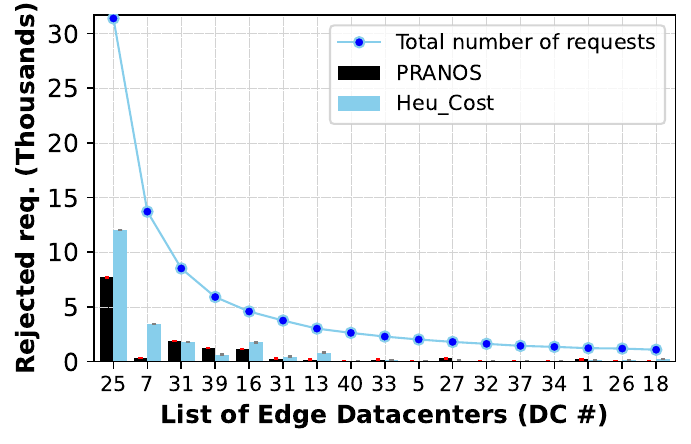}
    \caption{Rejected User Requests for 40N60E topology, for the scenario of Mixed Latency/Relaxed, Zipf (a=1.2), and 2 SFCs}
    \label{fig:detail}
\end{figure}

Figure~\ref{fig:40}(c), Figure~\ref{fig:100}(c), Figure~\ref{fig:citta}(c), Figure~\ref{fig:5gen}(c) show the number of rejected user requests under the hotspot scenario with relaxed latency constraints, but two different SFC topologies. Figure~\ref{fig:40}(d), Figure~\ref{fig:100}(d), Figure~\ref{fig:citta}(d) perform the same experiment, but with mixed latency constraints, meaning that one SFC topology has relaxed latency constraints and the second one has strict ones. It can be readily observed that in this scenario, as before, the global optimization of \ref{alg:prano} has fewer options to exploit, but it still outperforms HEU\_Cost on smaller topologies and is very close to the fractional lower bound $\fraqopt$. In Figure~\ref{fig:detail}, one can see a typical case of how user rejections are distributed across different \dcs in the physical substrate. The \dcs are ordered from most loaded to least loaded. As one can see, \ref{alg:prano} is indeed most effective in alleviating user rejections in the hotspot \dcs.

\subsubsection*{Cost of SFC Deployment}
In Figure~\ref{fig:40}(f), Figure~\ref{fig:100}(f), Figure~\ref{fig:citta}(f), Figure~\ref{fig:5gen}(f) and Figure~\ref{fig:40}(g), Figure~\ref{fig:100}(g), Figure~\ref{fig:citta}(g), Figure~\ref{fig:5gen}(g), we show how ECUs are distributed across the physical topology layers in the hotspot \dc scenario, in relaxed and mixed latency restrictions, respectively. ~\ref{alg:prano} succeeds to allocate much more ECUs in the cheaper layers of the physical topology that HEU\_Cost under all conditions.

\ignore{
\Cref{fig:utilization_uniform_150} shows how ECUs are distributed across RAN, Transport, and Core. This affects the overall cost of a solution. As \cref{fig:utilization_uniform_150} shows in the \emph{Strict} and \emph{Mixed} test cases, the direct comparison of the overall cost is not possible, because our algorithm accepts considerably more user sessions. However, if one fixes the number of accepted user sessions at the maximum achieved by SOTA, then, as one can see, Fluid uses much more \ecus in the Core, which makes the solution much cheaper. Thus, the proposed model is cost-efficient.}

\subsubsection*{Execution Time}

\begin{figure*}
    \captionsetup[subfloat]{belowskip=-1pt,font=scriptsize}
    \setkeys{Gin}{width=.25\linewidth,height=2.4cm,trim=0 .5cm 0 0,clip}
    \subfloat[Random Graph (40N 60E)\label{fig:time40}]{\includegraphics{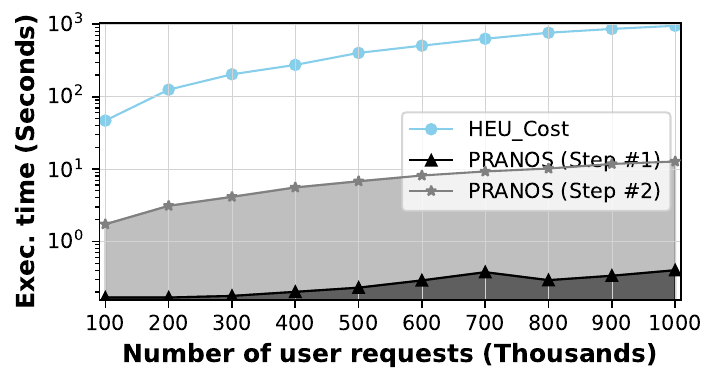}}
    \hfill
    \subfloat[Random Graph (100N 150E)\label{fig:time100}]{\includegraphics{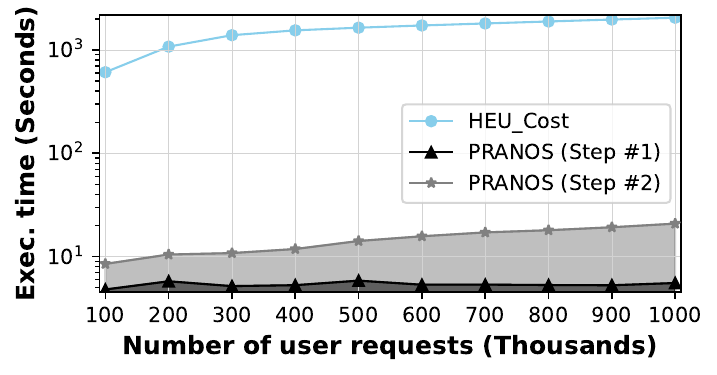}}
    \hfill
    \subfloat[Citta Studi (30N 35E)\label{fig:timecita}]{\includegraphics{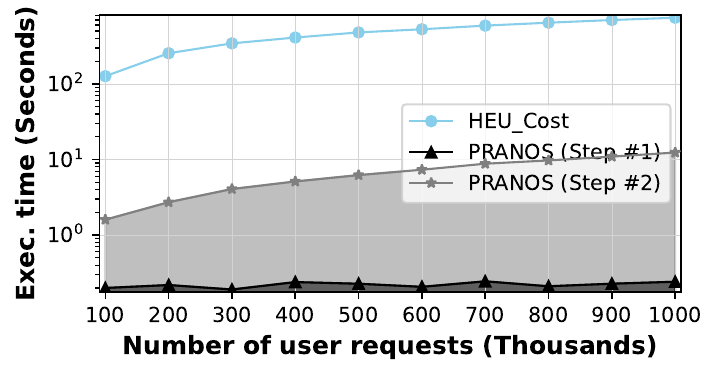}}
    \hfill
    \subfloat[5GEN (78N 100E)\label{fig:timegen}]{\includegraphics{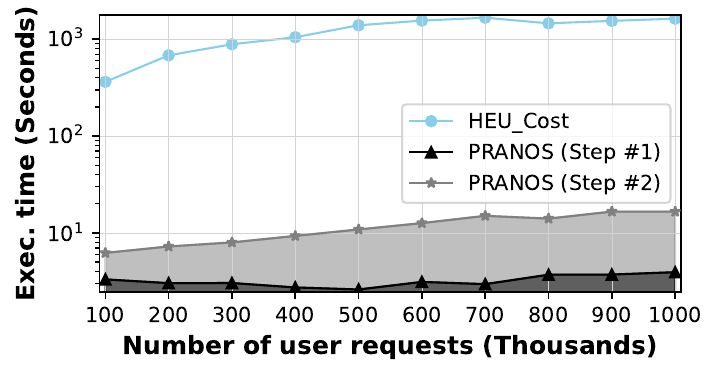}}
    \caption{Execution Time\label{fig:time}}
\end{figure*}

In Figure~\ref{fig:time}, we show the execution time as a function of user requests for selected physical topologies on the logarithmic scale. As expected, the execution time for the function placement and routing step of \ref{alg:prano} remains flat, independent of the number of user requests. The execution time of user allocation and request steering, the second step of \ref{alg:prano}, grows linearly with the number of users, but it grows considerably slower than that of HEU\_Cost. The reason is that even though \ref{alg:prano}'s user allocation phase is greedy, its search space is much more limited, because any path it finds in the fractional optimum $\fraqopt$, is a feasible path. Therefore, there is no need to backtrack and recalculate the shortest paths after each user request allocation. 

\ignore{
\Cref{fig:timeperuser} shows the execution time of the planning and allocation phases of our algorithm as a function of the number of users. We could not run SOTA fast enough beyond $150K$ users. Thus, we only report the results of the Fluid model. 
\ignore{In each experiment the average load on the system is maintained constant by matching the link and \dc capacity to the overall demand, but no \dcs are added. }

As expected, the execution time of the planning phase remains flat. The reason is that our model aggregates individual users by their access \dcs so the number of decision variables does not grow. 
This is critical for the practicality of the proposed approach: because planning can be carried out so fast even for a large topology, re-optimization, and plan corrections are cheap. Furthermore, the short planning time gets amortized quickly with the system's continuous operation.

The computational time of the second stage of our solution grows linearly with the number of users, however, is faster than SOTA. Even though our online user allocation phase is greedy, its search space is much more limited, because any path it finds on the output of the planning phase, is a feasible path. There is no need to backtrack and to recalculate shortest paths after each allocation. }

\graphicspath{{fig/INFOCOM24/}}
\setkeys{Gin}{width=0.25\textwidth}
\DeclareGraphicsExtensions{pdf}

\def\saturation{saturation}
\definecolor{light-gray}{gray}{0.9}
\def\netgen{5GEN}
\def\netA{10N20E}
\def\netB{20N30E}
\def\netC{40N60E}
\def\netD{50N50E}
\def\netE{60N90E}
\def\netF{80N120E}
\def\netG{100N150E}
\def\netstudi{cittastudi}

\def\netggen{5GEN}
\def\netgA{10N}
\def\netgB{20N}
\def\netgC{40N}
\def\netgD{50N}
\def\netgE{60N}
\def\netgF{80N}
\def\netgG{100N}
\def\netgstudi{Cittastudi}



\def\runA{RelaxedUniform1App}
\def\runB{RelaxedUniform2App}
\def\runC{RelaxedZipf1App}
\def\runD{RelaxedZipf2App}
\def\runE{StrictZipf1App}
\def\runF{StrictZipf1App(800)}
\def\runG{StrictZipf1App1200_}
\def\runH{MixedUniform2App}
\def\runI{MixedZipf2App}
\def\runTIME{execTime}

\def\runtA{{Relaxed\\Uniform\\1 App}}
\def\runtB{{Relaxed\\Uniform\\2 App}}
\def\runtC{{Relaxed\\Zipf\\1 App}}
\def\runtD{{Relaxed\\Zipf\\2 App}}
\def\runtE{{Strict\\Zipf\\1 App}}
\def\runtF{{Strict $(0.8)$$\\Zipf\\1 App}}
\def\runtG{{Strict\\Zipf\\1 App}}
\def\runtH{Mixed\\Uniform\\2 App}
\def\runtI{Mixed\\Zipf\\2 App}

\def\hhh{2.6cm}
\newlength{\vvv}
\newlength{\yyy}
\newlength{\zzz}\setlength{\zzz}{-5pt}
\newlength{\zzzz}\setlength{\zzzz}{-8pt}
\setlength{\vvv}{.4\baselineskip}
\setlength{\yyy}{.4\baselineskip}

\newcommand{\mysubfloat}[2][]{\rule{0pt}{\hhh}\includegraphics[#1]{#2}\\[\vvv]}

\newcommand{\gmysubfloat}[3][\hhh]{\rule{0pt}{#1}\includegraphics[#3]{#2}\\}

\newcommand{\tmysubfloat}[4][\yyy]{\parbox[b][#2][t]{.5\linewidth}{\centering\subfloat[]{~}\\[.5\baselineskip]#4}\hfill\rule{0pt}{#2}\parbox[b][#2][c]{\baselineskip}{\rotatebox[origin=lt]{90}{\tiny\textbf{#3}}}\\[#1]}

\newcommand{\gtmysubfloat}[4][\yyy]{\parbox[b][#2][t]{.5\linewidth}{\centering\subfloat[]{~}\\[.5\baselineskip]\textbf{#3}\\#4}\hfill\rule{0pt}{#2}\parbox[b][#2][c]{\baselineskip}{\rotatebox[origin=lt]{90}{~}}\\[#1]}

\def\doit(#1,#2,#3,#4,#5,#6){
\fboxsep1pt
\fcolorbox{#5}{white}%
{\begin{minipage}[t]{0.22\linewidth}
    \centering
    \setkeys{Gin}{width=\linewidth,height=\hhh,trim=0 .5cm 0 0,clip}
    \mysubfloat{saturation\runC#1.pdf}
    \mysubfloat{saturation\runG#1.pdf}
    \mysubfloat{saturation\runD#1.pdf}
    \mysubfloat{saturation\runI#1.pdf}
    \mysubfloat{saturation\runH#1.pdf}
    \mysubfloat{dc_share_utilization\runC#1.pdf}
    \mysubfloat{dc_share_utilization\runI#1.pdf}
    \gmysubfloat[3cm]{#2.pdf}{#4}   
    \caption{#3}\label{#6}
\end{minipage}\hspace{0pt}}
}

\begin{figure*}
    \centering
    \begin{minipage}[t]{0.075\linewidth}
        \centering\scriptsize
        \captionsetup[subfloat]{belowskip=-1pt,font=scriptsize}
        \addtolength{\yyy}{0pt}
        \tmysubfloat{\hhh}{Rejected requests (Thousands)}{\runtC}
        \tmysubfloat{\hhh}{Rejected requests (Thousands)}{\runtG}
        \tmysubfloat{\hhh}{Rejected requests (Thousands)}{\runtD}
        \tmysubfloat{\hhh}{Rejected requests (Thousands)}{\runtI}
        \tmysubfloat{\hhh}{Rejected requests (Thousands)}{\runtH}
        \tmysubfloat{\hhh}{Allocated \ecus (Millions)}{\runtC}
        \tmysubfloat{\hhh}{Allocated \ecus (Millions)}{\runtI}
        \gtmysubfloat[1cm]{3cm}{network topology}{}
    \end{minipage}\hspace{\zzzz}
    \doit(\netC,\netgC,\netC,{height=3cm,trim=0.5cm 5cm 0.5cm 6cm,clip},light-gray,fig:40)\hspace{\zzz}
    \doit(\netG,\netgG,\netG,{height=3cm,trim=0.5cm 5cm 0.5cm 5cm,clip},light-gray,fig:100)\hspace{\zzz}
    \doit(\netstudi,\netgstudi,{Citt{\'a}~Studi},{height=3cm,trim=0.5cm 6cm 0.5cm 6cm,clip},light-gray,fig:citta)\hspace{\zzz}
    \doit(\netgen,\netggen,\netgen,{height=3cm,trim=0.5cm 7cm 0.5cm 6cm,clip},light-gray,fig:5gen)
\end{figure*}

\section{Related Work} \label{sec:related} 

Our solution is similar in spirit to~\cite{RostSchmid-RandomizedRounding-ToN2019} and ~\cite{MunkRostRackeSchmid-relax-2021}. We also observe that a single VNF-FG embedding request is small compared to the total capacity of the physical network substrate, and use Linear Program (LP) to obtain a fractional solution to the problem that is then rounded to embed individual requests. 

Without latency constraints, the problem is close to  the minimum cost Multi-Commodity Flow (MCF) problem~\cite{garg2007faster}). 
Variants of the problem that also consider latency constraints have been studied in~\cite{BehraveshHCR-TNSM2021, chowdhury2009virtual, harutyunyan2019latency, harutyunyan2020latency, rahman2010survivable, chochlidakis2016low} . An innate property of these solutions is that they attempt greedily to embed user requests one by one. While this approach is very reasonable when user load is uniformly distributed across edge data centers through which the users enter the network, the greedy approach becomes disadvantageous to a global optimization of \ref{alg:prano} in case of uneven load distribution.

While handling latency requirements is difficult in a general case, we exploit the fact that in practical physical network substrates latency can be inferred from the layers of topology (access, aggregation, core). For example, embedding a VNF-FG with pair-wise latency constraints within a typical 5G/6G topology, only requires to arbitrate among the three layers of the topology (RAN, Transport, and Core) for each logical link, based on how strict the latency requirement on this link is.  

Our approach to the traffic flow aggregation is close to~\cite{feng2017approximation}. However, we provide a practical solution to satisfy latency constraints and evaluate it in very large practical scenarios.

In~\cite{MaoSYY-INFOCOM2022}, joint resource management and flow scheduling for SFC deployment in hybrid edge-and-cloud network is studied and constant factor approximation on cost and latency are provided. While in the worst case, the approximation ratio is large, the authors show that in many realistic scenarios they achieve approximation ratio between $1$ and $2.375$ for cost and latency. Similarly to our work, that paper explores pair-wise latency constraints model. In contrast to our study,~\cite{MaoSYY-INFOCOM2022} considers only simple chain topology, does not explore request rejection rate and conducts a considerably smaller scale evaluation study. 

In~\cite{XiangEMN-2023}, a problem called Joint Planning and Slicing of mobile Network and edge Computation resources (JPSNC) is studied comprehensively. The goal of JPSNC is to minimize a weighted sum of the total latency and network operation cost for serving several types of application traffic under the constraints of application maximum tolerable latency and overall network planning budget. JPSNC is formulated as a nonlinear problem and then a new heuristic is provided by the authors that compares favorably to greedy heuristics and in some cases provides near-optimal results. In~\cite{XiangEMN-2023} one of the more compelling up to date evaluation studies is presented. Some differences in modeling (e.g., JPSNC does not consider request rejection rate) prevent direct comparison between PRANOS and~\cite{XiangEMN-2023}, we use network topologies discussed in this paper as part of our own evaluation study and also compare PRANOS to theoretical optimum to properly position it. 

\ignore{
SFC deployment problem was studied also for the online setting. In  an online setting, requests arrive one by one. In the latter case, VNF migration should be minimized to optimize the total value of embedding~\cite{BehraveshHCR-TNSM2021,ChowdhuryRB-TON2012}. In the future, we plan to explore a method, in which the first sub-problem, VNF-FG embedding, is done based on a forecast and serves as a plan that informs individual requests mapping one by one. Provided that the forecast is high fidelity one, statistical multiplexing should result in minimization of VNF migration. 
}
\ignore{
Third, if our aggregated LP used to represent aggregated sustained demand, the sustained aggregate traffic remains the same, which alleviates the problem of migrations in the online setting. Many forecasting techniques exist. In this paper, we leave the forecasting problem out of scope. 
}

Recently, meta-heuristic approaches to SFC embedding have attracted considerable attention
~\cite{Ruiz2018, Cao2017, Kiran2021}. Another approach that rapidly becomes popular is applying deep reinforcement learning techniques~\cite{NFVdeep-2019, DeepRL-2021}. Improving these method (e.g., faster convergence, higher fidelity, handling of local minima) is an active field of investigation.

\section{Conclusions and Future Work}\label{sec:conclusions}

We presented~\ref{alg:prano}, a novel highly scalable heuristic for offline SFC deployment problem and studied its performance via extensive large-scale simulations. We show that for large realistic network topologies and hundreds of thousands of requests,~\ref{alg:prano} is superior to the state-of-the-art heuristics and comes very close to the theoretical lower bound in terms of requests rejection ratio. We show that the execution time of~\ref{alg:prano} grows slowly with the size of the problem.

Our future directions include exploring online setting and extending~\ref{alg:prano} to deal with more general topologies using tree decomposition and researching the utility of techniques like multi-path
TCP/IP~\cite{peng2013low} to overcome non-splittability constraint of a single request and therefore further reduce the user rejection rate. 

\ignore{
Another future direction is to apply~\pranos~to the online or a semi-online setting. Namely, a forecast of sustained request demand for point of presence data centers can be used as input for the first stage of~\pranos~that would build a blueprint of a VNF-FG graph embedding that will inform one by one processing of individual requests or sets of requests as they arrive into the network. This way, thanks to statistical multiplexing we expect to gain a very low VNF migration rate.}

\cleardoublepage
{
\IEEEtriggeratref{18}
\bibliographystyle{IEEEtran}
\bibliography{IEEEabrv,main}

\begin{thebibliography}{10}
\providecommand{\url}[1]{#1}
\csname url@samestyle\endcsname
\providecommand{\newblock}{\relax}
\providecommand{\bibinfo}[2]{#2}
\providecommand{\BIBentrySTDinterwordspacing}{\spaceskip=0pt\relax}
\providecommand{\BIBentryALTinterwordstretchfactor}{4}
\providecommand{\BIBentryALTinterwordspacing}{\spaceskip=\fontdimen2\font plus
\BIBentryALTinterwordstretchfactor\fontdimen3\font minus \fontdimen4\font\relax}
\providecommand{\BIBforeignlanguage}[2]{{%
\expandafter\ifx\csname l@#1\endcsname\relax
\typeout{** WARNING: IEEEtran.bst: No hyphenation pattern has been}%
\typeout{** loaded for the language `#1'. Using the pattern for}%
\typeout{** the default language instead.}%
\else
\language=\csname l@#1\endcsname
\fi
#2}}
\providecommand{\BIBdecl}{\relax}
\BIBdecl

\bibitem{KAUR2020100298}
K.~Kaur, V.~Mangat, and K.~Kumar, ``{A Comprehensive Survey of Service Function Chain Provisioning Approaches in SDN and NFV Architecture},'' \emph{{Computer Science Review}}, vol.~38, p. 100298, 2020.

\bibitem{Rexford2008}
M.~Yu, Y.~Yi, J.~Rexford, and M.~Chiang, ``{Rethinking Virtual Network Embedding: Substrate Support for Path Splitting and Migration},'' \emph{SIGCOMM Comput. Commun. Rev.}, vol.~38, no.~2, p. 17–29, 2008.

\bibitem{RostSchmidt-ToN2020}
M.~Rost and S.~Schmid, ``{On the Hardness and Inapproximability of Virtual Network Embeddings},'' \emph{IEEE/ACM Transactions on Networking}, vol.~28, no.~2, pp. 791--803, 2020.

\bibitem{EvenIS-Multicommodity-Unsplittable-Flow-1975}
S.~Even, A.~Itai, and A.~Shamir, ``{On the Complexity of Time Table and Multi-commodity Flow Problems},'' in \emph{16th Annual Symposium on Foundations of Computer Science}, 1975, pp. 184--193.

\bibitem{Liu-5GKPI-2020}
{Guangyi Liu and Dajie Jiang}, ``{5G: Vision and Requirements for Mobile Communication System towards Year 2020},'' \emph{{Chinese Journal of Engineering}}, 2020.

\bibitem{OughtonFGB-Netherlands5G-2019}
{Oughton, E, Z Frias, S Van Der Gaast, and R Van Der Berg}, ``{Assessing the Capacity, Coverage and Cost of 5G Infrastructure Strategies: Analysis of The Netherlands},'' \emph{{Telematics and Informatics}}, vol.~37, p. 50–69, 2019.

\bibitem{RostSchmid-RandomizedRounding-ToN2019}
M.~Rost and S.~Schmid, ``{Virtual Network Embedding Approximations: Leveraging Randomized Rounding},'' \emph{IEEE/ACM Transactions on Networking}, vol.~27, no.~5, pp. 2071--2084, 2019.

\bibitem{MunkRostRackeSchmid-relax-2021}
R.~Münk, M.~Rost, H.~Räcke, and S.~Schmid, ``{It's Good to Relax: Fast Profit Approximation for Virtual Networks with Latency Constraints},'' in \emph{IFIP Networking Conference}, 2021, pp. 1--3.

\bibitem{harutyunyan2019latency}
D.~Harutyunyan, N.~Shahriar, R.~Boutaba, and R.~Riggio, ``{Latency-aware Service Function Chain Placement in 5G Mobile Networks},'' in \emph{IEEE Conference on Network Softwarization (NetSoft)}.\hskip 1em plus 0.5em minus 0.4em\relax IEEE, 2019, pp. 133--141.

\bibitem{harutyunyan2020latency}
------, ``{Latency and Mobility-aware Service Function Chain Placement in 5G Networks},'' \emph{IEEE Transactions on Mobile Computing}, 2020.

\bibitem{ChowdhuryRB-TON2012}
M.~Chowdhury, M.~R. Rahman, and R.~Boutaba, ``{ViNEYard: Virtual Network Embedding Algorithms With Coordinated Node and Link Mapping},'' \emph{IEEE/ACM Transactions on Networking}, vol.~20, no.~1, pp. 206--219, 2012.

\bibitem{SAP-HANA}
``{SAP HANA Cloud Capacity Unit Estimator},'' \url{ttps://hcsizingestimator.cfapps.eu10.hana.ondemand.com}, 2022.

\bibitem{RostDohneSchmid-parameterized-2019}
M.~Rost, E.~D\"{o}hne, and S.~Schmid, ``{Parametrized Complexity of Virtual Network Embeddings: Dynamic \& Linear Programming Approximations},'' \emph{SIGCOMM Comput. Commun. Rev.}, vol.~49, no.~1, p. 3–10, feb 2019.

\bibitem{MaoSYY-INFOCOM2022}
Y.~Mao, X.~Shang, and Y.~Yang, ``{Joint Resource Management and Flow Scheduling for SFC Deployment in Hybrid Edge-and-Cloud Network},'' in \emph{IEEE INFOCOM}, 2022, pp. 170--179.

\bibitem{BehraveshHCR-TNSM2021}
R.~Behravesh, D.~Harutyunyan, E.~Coronado, and R.~Riggio, ``{Time-Sensitive Mobile User Association and SFC Placement in MEC-Enabled 5G Networks},'' \emph{IEEE Transactions on Network and Service Management}, vol.~18, no.~3, pp. 3006--3020, 2021.

\bibitem{gurobi}
\BIBentryALTinterwordspacing
``{Gurobi Mathematical Optimization Solver},'' {Accessed on 20.03.2022}. [Online]. Available: \url{https://www.gurobi.com/}
\BIBentrySTDinterwordspacing

\bibitem{Karakostas-fractionalMCF2008}
\BIBentryALTinterwordspacing
G.~Karakostas, ``{Faster Approximation Schemes for Fractional Multicommodity Flow Problems},'' \emph{ACM Trans. Algorithms}, vol.~4, no.~1, mar 2008. [Online]. Available: \url{https://doi.org/10.1145/1328911.1328924}
\BIBentrySTDinterwordspacing

\bibitem{SCHARDONG2021107726}
\BIBentryALTinterwordspacing
F.~Schardong, I.~Nunes, and A.~Schaeffer-Filho, ``{NFV Resource Allocation: a Systematic Review and Taxonomy of VNF Forwarding Graph Embedding},'' \emph{Computer Networks}, vol. 185, p. 107726, 2021. [Online]. Available: \url{https://www.sciencedirect.com/science/article/pii/S1389128620313189}
\BIBentrySTDinterwordspacing

\bibitem{XiangEMN-2023}
B.~Xiang, J.~Elias, F.~Martignon, and E.~Nitto, ``{Joint Planning of Network Slicing and Mobile Edge Computing: Models and Algorithms},'' \emph{IEEE Transactions on Cloud Computing}, vol.~11, no.~01, pp. 620--638, jan 2023.

\bibitem{5GEN}
J.~Martín-Pérez, L.~Cominardi, C.~J. Bernardos, and A.~Mourad, ``{5GEN: A tool to generate 5G infrastructure graphs},'' in \emph{IEEE Conference on Standards for Communications and Networking (CSCN)}, 2019, pp. 1--4.

\bibitem{Erdos:1959:pmd}
P.~Erd\"os and A.~R\'enyi, ``On random graphs i,'' \emph{Publicationes Mathematicae Debrecen}, vol.~6, pp. 290--297, 1959.

\bibitem{Sharma2021}
S.~Sharma, A.~Engelmann, A.~Jukan, and A.~Gumaste, ``{VNF Availability and SFC Sizing Model for Service Provider Networks},'' \emph{IEEE Access}, vol.~8, pp. 119\,768--119\,784, 2020.

\bibitem{garg2007faster}
N.~Garg and J.~K{\"o}nemann, ``Faster and simpler algorithms for multicommodity flow and other fractional packing problems,'' \emph{SIAM Journal on Computing}, vol.~37, no.~2, pp. 630--652, 2007.

\bibitem{chowdhury2009virtual}
N.~M.~K. Chowdhury, M.~R. Rahman, and R.~Boutaba, ``{Virtual Network Embedding with Xoordinated Node and Link Mapping},'' in \emph{IEEE INFOCOM}.\hskip 1em plus 0.5em minus 0.4em\relax IEEE, 2009, pp. 783--791.

\bibitem{rahman2010survivable}
M.~R. Rahman, I.~Aib, and R.~Boutaba, ``{Survivable Virtual Network Embedding},'' in \emph{International Conference on Research in Networking}.\hskip 1em plus 0.5em minus 0.4em\relax Springer, 2010, pp. 40--52.

\bibitem{chochlidakis2016low}
G.~Chochlidakis and V.~Friderikos, ``{Low Latency Virtual Network Embedding for Mobile Networks},'' in \emph{IEEE International Conference on Communications (ICC)}.\hskip 1em plus 0.5em minus 0.4em\relax IEEE, 2016, pp. 1--6.

\bibitem{feng2017approximation}
H.~Feng, J.~Llorca, A.~M. Tulino, D.~Raz, and A.~F. Molisch, ``{Approximation Algorithms for the NFV Service Distribution Problem},'' in \emph{IEEE INFOCOM}.\hskip 1em plus 0.5em minus 0.4em\relax IEEE, 2017, pp. 1--9.

\bibitem{Ruiz2018}
L.~Ruiz, R.~J. Dur{\'{a}}n, I.~de~Miguel, P.~S. Khodashenas, J.~J. Pedreno-Manresa, N.~Merayo, J.~C. Aguado, P.~Pavon-Marino, S.~Siddiqui, J.~Mata, P.~Fern{\'{a}}ndez, R.~M. Lorenzo, and E.~J. Abril, ``{A Genetic Algorithm for VNF Provisioning in NFV-enabled Cloud/MEC RAN Architectures},'' \emph{Applied Sciences (Switzerland)}, vol.~8, no.~12, 2018.

\bibitem{Cao2017}
J.~Cao, Y.~Zhang, W.~An, X.~Chen, J.~Sun, and Y.~Han, ``{VNF-FG Design and VNF Placement for 5G Mobile Networks},'' \emph{Science China Information Sciences}, vol.~60, no.~4, pp. 1--15, 2017.

\bibitem{Kiran2021}
N.~Kiran, X.~Liu, S.~Wang, and C.~Yin, ``{Optimising resource allocation for virtual network functions in SDN/NFV-enabled MEC networks},'' \emph{IET Communications}, vol.~15, no.~13, pp. 1710--1722, 2021.

\bibitem{NFVdeep-2019}
X.~Yikai, Z.~Qixia, L.~Fangming, W.~Jia, Z.~Miao, Z.~Zhongxing, and Z.~Jiaxing, ``{NFVDeep: Adaptive Online Service Function Chain Deployment with Deep Reinforcement Learning},'' in \emph{{IWQoS '19: Proceedings of the International Symposium on Quality of Service}}, 2019.

\bibitem{DeepRL-2021}
Y.~Liu, Y.~Lu, X.~Li, W.~Qiao, Z.~Li, and D.~Zhao, ``{SFC Embedding Meets Machine Learning: Deep Reinforcement Learning Approaches},'' \emph{{IEEE Communications Letters}}, vol.~25, no.~6, pp. 1926--1930, 2021.

\bibitem{peng2013low}
Q.~Peng, A.~Walid, and H.~Steven, ``{Multipath TCP Algorithms: Theory and Design},'' in \emph{ACM International Conference on Measurement and Modeling of Computer Systems (SIGMETRICS)}, vol.~13, 2013, pp. 17--21.

\end{thebibliography}
}
\listoffixmes
\newpage




\end{document}